\documentclass[aps,prd,reprint,twocolumn,10pt,eqsecnum,showpacs,nofootinbib,amsfonts,amssymb,amsmath,longbibliography,superscriptaddress,floatfix]{revtex4-1}
\usepackage[colorlinks=true,allcolors=blue]{hyperref}
\usepackage{bm}
\usepackage{mathrsfs}
\usepackage{xcolor}
\usepackage{mathtools}
\usepackage{graphicx}
\usepackage{multirow}

\DeclareSymbolFont{bbold}{U}{bbold}{m}{n}
\DeclareSymbolFontAlphabet{\mathbbold}{bbold}

\newcommand{\ud}{\mathrm{d}}
\newcommand{\pb}[1]{\,\mbox{}_{#1}}
\newcommand{\mc}[1]{\mathcal{#1}}
\newcommand{\ms}[1]{\mathscr{#1}}
\newcommand{\bs}[1]{\boldsymbol{#1}}

\renewcommand{\>}{\rangle}
\newcommand{\interchange}[2]{#1 \longleftrightarrow #2}
\renewcommand{\Re}{\operatorname{Re}}
\renewcommand{\Im}{\operatorname{Im}}
\newcommand*{\ditto}{\texttt{"}}
\newcommand{\expect}{\mathbb{E}}

\DeclareMathOperator{\erf}{erf}
\DeclareMathOperator{\sech}{sech}

\usepackage[normalem]{ulem}

\begin{document}

\author{Alexander M.\ Grant}
\email{alex.grant@virginia.edu}

\author{David A.\ Nichols}
\email{david.nichols@virginia.edu}

\affiliation{Department of Physics, University of Virginia, P.O.~Box 400714,
  Charlottesville, Virginia 22904-4714, USA}

\title{Outlook for detecting the gravitational wave displacement and spin memory effects with current and future gravitational wave detectors}

\begin{abstract}
  Gravitational wave memory effects arise from non-oscillatory components of gravitational wave signals, and they are predictions of general relativity in the nonlinear regime that have close connections to the asymptotic properties of isolated gravitating systems.
  There are many types of memory effects that have been studied in the literature.
  In this paper we focus on the ``displacement'' and ``spin'' memories, which are expected to be the largest of these effects from sources such as the binary black hole mergers which have already been detected by LIGO and Virgo.
  The displacement memory is a change in the relative separation of two initially comoving observers due to a burst of gravitational waves, whereas the spin memory is a portion of the change in relative separation of observers with initial relative velocity.
  As both of these effects are small, LIGO, Virgo, and KAGRA can only detect memory effects from individual events that are much louder (and thus rarer) than those that have been detected so far.
  By combining data from multiple events, however, these effects could be detected in a population of binary mergers.
  In this paper, we present new forecasts for how long current and future detectors will need to operate in order to measure these effects from populations of binary black hole systems that are consistent with the populations inferred from the detections from LIGO and Virgo's first three observing runs.
  We find that a second-generation detector network of LIGO, Virgo, and KAGRA operating at the O4 (``design'') sensitivity for 1.5 years and then operating at the O5 (``plus'') sensitivity for an additional year can detect the displacement memory.
  For Cosmic Explorer, we find that displacement memory could be detected for individual loud events, and that the spin memory could be detected in a population within 2 years of observation time.
\end{abstract}

\maketitle

\tableofcontents

\section{Introduction}

Since the first detection of gravitational waves from a binary black hole merger by LIGO~\cite{LSC2016}, gravitational waves from nearly one-hundred binaries have now been detected in three observing runs (O1--O3)~\cite{LSC2018a, LSC2020a, LSC2021a}.
These detections enabled general relativity to be tested through many methods~\cite{LSC2019a, LSC2020b, LSC2021c}, and they provided constraints on the astrophysical populations of such binaries~\cite{LSC2018b, LSC2020c, LSC2021b}.
The gravitational-wave tests of general relativity are complementary to the constraints solar-system and pulsar measurements provided (see, for example,~\cite{Will2014} and references therein), because the radiation emitted by the merger of black holes probes into the dynamical and strong-field regime that was not accessible to these earlier measurements.
The gravitational-wave features which were measured and used in the tests of general relativity primarily arose from the dominant, quadrupolar waves.
However, some distinctive strong-field predictions of general relativity appear in subleading portions of the waveform.
It is then natural to ask what \emph{subdominant} phenomena can be measured by current (and future) gravitational wave detectors, and when will it be possible?

In this paper, we aim to address these questions for two nonlinear relativistic phenomena known as the displacement~\cite{Zeldovich1974} and spin~\cite{Pasterski2015} memory effects.
These effects can be determined by sets of observers who measure enduring changes in their separation before and after a burst of gravitational waves.
The displacement memory arises for observers who are initially comoving, and the spin memory (together with the related center-of-mass memory~\cite{Nichols2018}) arises for observers with initial relative velocities~\cite{Flanagan2019, Grant2021}.
For interferometric gravitational wave detectors (such as LIGO), which measure gravitational waves over a finite time with a limited frequency bandwidth, these effects are encompassed in nonoscillatory parts of the measured signal.
For the displacement memory, there is a distinctive part of the signal associated with the net change between early and late times, while the spin memory has an analogous portion related to the nonzero time integral of the signal.

Memory effects have close connections to the infrared properties of gravity and gauge theories, including the asymptotic field equations, symmetries, and conserved charges (see, for example,~\cite{Strominger:2017zoo}).
In particular, the displacement memory is related to the supertranslation symmetries~\cite{Strominger2014, Newman1966}, which are a subgroup of the Bondi-Metzner-Sachs (BMS) group of symmetries for asymptotically flat spacetimes~\cite{Bondi1962, Sachs1962b}, and they are closely connected with the conserved charges conjugate to these symmetries~\cite{Strominger2014, Flanagan2015}.
BMS symmetry and memory effects have implications for the quantization of gravity at null infinity~\cite{Ashtekar1987, Strominger2014, Prabhu2022} which are under active investigation.
In addition, there are proposed extensions to the BMS group~\cite{Barnich2009, Campiglia2014} which have conjugate charges that are related to the spin memory effect~\cite{Pasterski2015}.
The relationship between the spin memory and the generalized BMS symmetry is more involved, however~\cite{Compere:2018ylh}.
Because the spin memory effect can be described by and derived from the asymptotic Einstein equations without the use of these symmetries and their charges (see, for example,~\cite{Grant2021}), the effect exists independently of the particular proposal for the extension of the BMS symmetry algebra (and its measurement would not give evidence for or against a given proposal).

Memory signals from binary black hole mergers are also clear and distinctive probes of nonlinearities in Einstein's equations that are not as apparent in the oscillatory parts of the signal.
In the linear theory, the displacement memory effect vanishes for gravitationally bound systems (such as black hole binaries detected by LIGO), as it arises only for unbound bodies and fields (for example, in scattering~\cite{Zeldovich1974} and in the radiated neutrinos in supernovae~\cite{Epstein1978, Turner1978}).
There is, however, a non-negligible contribution in the nonlinear theory~\cite{Christodoulou1991}.
Thorne~\cite{Thorne1992} interpreted this as the nonlinear effective stress-energy tensor of the gravitational waves acting as the ``unbound'' material producing the memory.
Because it is sourced by the oscillatory waves, the displacement memory effect (and, similarly, the spin memory effect) probes nonlinearities in the \emph{propagation} of gravitational waves from an isolated source.

As binary black hole mergers are some of the most luminous sources in the universe (with luminosities approaching the Planck value), these systems enter the regime in which nonlinearities in the propagation of gravitational waves are important; they are thus well suited for producing the nonlinear displacement and spin memory effects.
For the sources observed by LIGO and Virgo to date, however, the signals were not sufficiently loud for there to be evidence for the memory effect in any of the individual detections~\cite{Hubner:2021amk}.
This is consistent with earlier forecasts~\cite{Favata2009, Johnson2018}.
Even with KAGRA~\cite{KAGRA2018} and LIGO India~\cite{Unnikrishnan2013} joining the network, our results show that it is unlikely that the memory effects will be detected from individual events even as the detectors reach their design and their ``plus'' sensitivities.

The outlook for detecting the memory from individual events is more promising with the space-based interferometer LISA~\cite{Amaro-Seoane2017}.
It is has been estimated that LISA will observe the displacement memory arising from mergers of supermassive black-hole binaries~\cite{Islo2019}.
Pulsar timing array experiments (see, for example,~\cite{Verbiest2021}) have put upper limits on the amplitude of gravitational-wave bursts with memory~\cite{NANOGrav2019}, though Ref.~\cite{Islo2019} also suggested that it will likely take longer for pulsar timing to detect the memory effect than LISA.
Finally, there are forecasts that show that next-generation ground-based detectors such as Einstein Telescope~\cite{Punturo2010} and Cosmic Explorer~\cite{Reitze2019} will be sensitive enough to measure the displacement memory from individual events~\cite{Johnson2018}.
We give further evidence for this in this paper.

While detecting the displacement memory effect from individual events is unlikely, it it is possible to measure the presence of the memory in the entire population of mergers by combining the evidence for the effect over all the observed events~\cite{Lasky2016}.
This procedure (often referred to as ``stacking'' and described in more detail in Sec.~\ref{sec:stacking} below) makes use of the fact that many low-significance events below the threshold of detection can be coherently combined to give a single higher-significance ``effective event'' that would exceed a threshold for detection.
Stacking has also been proposed as a means of measuring phenomena other than memory that are similarly small (for example, to search for features in the ringdown waves~\cite{Berti2018}).
In~\cite{Hubner2019,Hubner:2021amk}, this method was used to determine the statistical evidence for the presence of the displacement memory effect in the first two gravitational-wave transient catalogs.
There was not significant evidence for the memory in these events.
This is consistent with the results of~\cite{Lasky2016}, which determined that about 100 events similar to GW150914 would be necessary for detection of the memory in a population of events.

The study in~\cite{Lasky2016} was a proof-of-principle work, and the subsequent works~\cite{Hubner2019,Hubner:2021amk} developed a more complete detection pipeline and more accurate forecasts of the detection prospects.
For example, the analysis in~\cite{Hubner2019} used the events from the first gravitational-wave catalog and the associated astrophysical population models from the first two observing runs~\cite{LSC2018b}, and found that it would take $O(2000)$ events at the LIGO/Virgo design sensitivities to reach a detection threshold.
This would most likely occur during the fifth observing run (O5)~\cite{KAGRA2013}.
The methods used in~\cite{Hubner2019,Hubner:2021amk} were Bayesian, and they specifically involved computing an evidence ratio between the hypotheses that each event did or did not include the displacement memory.
These evidence-ratio calculations are computationally intensive, and having a faster method for performing forecasts is useful.
One such less computationally intensive approach was also discussed in~\cite{Lasky2016}.
The method is to add the signal-to-noise ratios of the memory part of each signal in quadrature---while excluding some lower-significance events where certain parameters cannot be adequately measured (which we discuss in detail in Sec.~\ref{sec:sign} below)---to compute an ``effective memory signal-to-noise ratio'' for the population of events.
This approach was performed in~\cite{Boersma2020}, which showed that a detection with an effective signal-to-noise ratio of 3 would be seen in five years of LIGO and Virgo operation at their design (O4) sensitivities.
The results of~\cite{Boersma2020} are consistent with those of~\cite{Hubner2019} in terms of the number of events which were needed.
In this paper, we adopt the approach of~\cite{Lasky2016, Boersma2020}, although we also illustrate in Sec.~\ref{sec:stacking} a limit in which it is equivalent to the Bayesian hypothesis test of~\cite{Hubner2019, Hubner:2021amk}.

The goals of this paper are twofold.
(i) We update the results of~\cite{Boersma2020} to use a population model that was informed by the second gravitational-wave catalog~\cite{LSC2020c} and to account for the updated observing scenarios of the current ground-based detectors.
Specifically, as outlined in~\cite{KAGRA2013}, we allow for the detector sensitivities to increase to those of O5 (the ``plus'' sensitivities) after $1.5$ years, with LIGO India added after $2.25$ years.
These two changes result in the effective signal-to-noise ratio for the memory reaching the threshold of 3 after about 2.5 years of total observation.
(ii) We perform a similar type of forecast for the detection prospects for the spin memory effect in a population of binary black hole mergers.
Previous estimates had suggested that the spin memory is about 10 times smaller than the displacement memory~\cite{Nichols2017}, which makes it too weak to be detected by the current generation of ground-based detectors, even in a population of events.
We thus perform the forecasts using next-generation detectors, and find that the effective signal-to-noise ratio of the spin memory could reach 3 in the Cosmic Explorer detector network (assuming three detectors) after 2 years of observation.
Furthermore, Cosmic Explorer can detect the displacement memory from individual loud events, as predicted in~\cite{Johnson2018}.

The outline of the remainder of the paper is as follows.
First, in Sec.~\ref{sec:detection}, we discuss various aspects of detecting the memory: the stacking procedure outlined above, the definition of the part of the waveform that gives rise to the memory effect, and an issue (related to the ``sign'' of a detector's response to this part of the waveform) that arises for certain events whose parameters are not well constrained.
In Sec.~\ref{sec:methods} we describe the methods we use for forecasting the memory: how one generates event parameters from a distribution of population parameters, the models we use to generate waveforms, and the details of how the signal-to-noise ratio is then computed for a given detector network.
We then give our results in Sec.~\ref{sec:results}, predicting the accumulated signal-to-noise ratio as a function of time, and we discuss a subtlety due to the current lack of constraints on the population of events at large redshift.
We give our conclusions and discuss future directions in Sec.~\ref{sec:conclusions}.

The code which we used to implement these forecasts, as well as compute the displacement and spin memories generally, is publically available online at~\cite{gwforecasts}.

\section{Formalism for detection of memory effects} \label{sec:detection}

In the three parts of this section, we discuss a few topics needed for our procedure to forecast the detection prospects of the memory.
These are the signal-to-noise ratio, the Bayes factor, and the ``stacking'' of events (Sec.~\ref{sec:stacking}); the memory signals used in the forecasts (Sec.~\ref{sec:mem_signal}); and a subtlety related to stacking associated with the ``sign'' of a detector's response to the memory effect's signal (Sec.~\ref{sec:sign}).

\subsection{Signal-to-noise ratio and stacking} \label{sec:stacking}

A common figure of merit used in assessing whether an event can be measured in a given detector (or set of detectors) is the so-called \emph{signal-to-noise ratio} (SNR).
In the context of gravitational waves, the SNR is commonly defined in the frequency domain.
To give the expression for the SNR, it is first helpful to define the noise-weighted inner product of two (real) signals $a(t)$ and $b(t)$ by
\begin{equation}
  \<a|b\> \equiv 2 \int_{-\infty}^\infty \frac{\tilde a(f) \overline{\tilde b (f)} \ud f}{S_n (f)} .
\end{equation}
The tildes above denote the Fourier transforms of $a$ and $b$, and $S_n (f)$ is the power spectral density of the noise, a quantity which characterizes the (assumed) stationary Gaussian noise in the detector (see, for example, Chapter~2 of~\cite{Stratonovich1963}).
For $a$ and $b$ real and $S_n(-f) = S_n(f)$, the inner product $\<a|b\>$ is also real.
Then, the optimal SNR $\rho_h$ associated with a given signal $h(t)$ is given by
\begin{equation}
  \rho_h \equiv \sqrt{\<h|h\>}.
\end{equation}

In~\cite{Lasky2016}, two approaches were used to forecast the detection prospects for the displacement memory effect.
The first directly involved the SNR.
More specifically, an effective SNR was computed by adding in quadrature the individual SNRs of the memory part of the signal in all detectors for all events.
The second method did not directly compute an SNR; rather, it performed Bayesian model comparison by computing (the log of) an evidence ratio for the hypotheses that the memory effect is present versus absent in the simulated events (this will be described more quantitatively below).
We now summarize the context in which the two approaches become equivalent to one another.

We start by reviewing the evidence-ratio calculation, which in the context of gravitational-wave memory detection, aims to answer the following question: given a set of data $d(t)$, which model is more favored: a waveform $h_{\rm osc.} (t)$ that has a vanishing memory effect, or $h_{\rm osc.} (t) + h_{\rm mem.} (t)$,\footnote{Note that, in this section, we do not define what ``$h_{\rm mem.} (t)$'' is, other than to distinguish between a model with versus without memory.
  See Sec.~\ref{sec:mem_signal} for a further discussion of the subtleties in defining this quantity, which we call the ``memory signal.''} which has a nonvanishing memory effect?
The probability that a set of data $d(t)$ is given by $s(t) + n(t)$, with $n(t)$ being some realization of stationary, Gaussian noise characterized by the power spectral density $S_n (f)$, is given by (see, for example,~\cite{Finn1992, Thrane2018}):
\begin{equation}
  \mc L[d | s] \propto \exp\left(-\frac{1}{2} \rho_{d - s}^2\right).
\end{equation}
In the language of Bayesian statistics, this is a \emph{likelihood}.\footnote{In the gravitational-wave literature, this is often written as $\mc L[d | \theta]$, which implicitly assumes some signal model $h(\theta)$; this will be written as $\mc L[d | h(\theta)]$ in our notation.}
Assuming that each model is equally favored beforehand, the Bayes factor, which assesses by how much the model with memory is favored over the model without, is given by
\begin{equation} \label{eqn:bayes_factor}
  \mc B^{\rm mem.}_{\rm no\,mem.} (d) = \frac{\displaystyle\int \ud \theta \mc L[d | h_{\rm osc.} (\theta) + h_{\rm mem.} (\theta)] p(\theta)}{\displaystyle \int \ud \theta \mc L[d | h_{\rm osc.} (\theta)] p(\theta)},
\end{equation}
where $p(\theta)$ is the (prior) probability that the parameters take on a particular value $\theta$.
Bayes' theorem,
\begin{equation}
  p[h(\theta) | d] p(d) = \mathcal L[d | h(\theta)] p(\theta),
\end{equation}
where $p(d)$ is the evidence for the model $h(\theta)$ and $p[h(\theta) | d]$ is the posterior probability of the parameters $\theta$ given the data $d$, can be used to rewrite the Bayes factor in Eq.~\eqref{eqn:bayes_factor}.
First, one can see that the Bayes factor is precisely the ratio of the evidences $p(d)$ of the two signal models.
Second, one can rewrite the Bayes factor in Eq.~\eqref{eqn:bayes_factor} in the form
\begin{equation} \label{eqn:bayes_factor_post}
  \mc B^{\rm mem.}_{\rm no\,mem.} (d) = \int \ud \theta \frac{\mathcal L[d | h_\mathrm{osc.}(\theta) + h_\mathrm{mem.}(\theta)]}{\mathcal L[d | h_\mathrm{osc.}(\theta)]}  p[h_\mathrm{osc.}(\theta) | d],
\end{equation}
where
\begin{equation}
  \begin{split}
    &\frac{\mc L[d | h_{\rm osc.} (\theta) + h_{\rm mem.} (\theta)]}{\mc L[d | h_{\rm osc.} (\theta)]} \\
    &\hspace{1em}= \exp\left[-\frac{1}{2} \rho_{d - h_{\rm osc.} (\theta) - h_{\rm mem.} (\theta)}^2 + \frac{1}{2} \rho_{d - h_{\rm osc.} (\theta)}^2\right]
  \end{split}
\end{equation}
is the ratio of the likelihoods of the two models and $p[h_{\rm osc.} (\theta) | d]$ is the posterior for the parameters $\theta$ under the assumption of a model containing just the oscillatory signal and no memory signal.
This posterior probability is the one that is most commonly computed for the observed events in the current gravitational-wave transient catalogs~\cite{LSC2020a, LSC2021a}.

Next, we make two assumptions.
First, we assume that the signal contains the memory effect, so that the set of data $d$ is given by
\begin{equation}
  d = h_{\rm osc.} (\theta_0) + h_{\rm mem.} (\theta_0) + n,
\end{equation}
where $n$ is a realization of the noise and $\theta_0$ are some ``true'' values of these parameters.
The second assumption is that we can approximate the posterior distribution by a delta function: namely,
\begin{equation}
  p[h_{\rm osc.} (\theta) | d] = \delta(\theta - \theta_0).
\end{equation}
In this approximation, we are neglecting the spread in $p[h_{\rm osc.} (\theta) | d]$ and biases from the true parameters that would arise from a particular realization of the detector's noise, as well as any errors to the parameter estimation that would arise from neglecting the memory part of the signal.
However, we focus on this approximation because it results in an exact relationship between the two data-analysis methods.
Note that, for reasons discussed in Sec.~\ref{sec:sign}, this is not always a good approximation, as $p[h_{\rm osc.} (\theta) | d]$ can have multiple peaks (and would not be well represented by a single delta function).

It then follows that the Bayes factor is given by
\begin{equation} \label{eqn:Bayes_with_noise}
  \begin{split}
    \mc B^{\rm mem.}_{\rm no\,mem.} (d) &= \exp\left[-\frac{1}{2} \rho_n^2 + \frac{1}{2} \rho_{h_{\rm mem.} (\theta_0) + n}^2\right] \\
    &= \exp\left[\frac{1}{2} \rho_{h_{\rm mem.} (\theta_0)}^2 + \<h_{\rm mem.} (\theta_0)|n\>\right].
  \end{split}
\end{equation}
We now average Eq.~\eqref{eqn:Bayes_with_noise} over the noise, which we denote by $\expect$ (for ``expectation value'').
We apply the result in Eq.~\eqref{eqn:avg_exp} to determine that
\begin{equation}
  \expect\left\{\exp\left[\<h_{\rm mem.} (\theta_0)|n\>\right]\right\} = \exp\left[\frac{1}{2} \rho_{h_{\rm mem.} (\theta_0)}^2\right],
\end{equation}
and then take the logarithm of the expectation value to obtain
\begin{equation} \label{eqn:logBayes}
  \ln \expect[\mc B^{\rm mem.}_{\rm no\,mem.} (d)] = \rho_{h_{\rm mem.} (\theta_0)}^2.
\end{equation}

The total Bayes factor for independent events is simply the product of the individual Bayes factors for each individual event.
Since the square of the SNR is the log of the Bayes factor in this approximation, it is apparent that the ``effective'' SNR squared of a series of independent events, as well as a single event measured in multiple detectors, is given simply by the sum:
\begin{equation} \label{eqn:stacking}
  \rho_{\rm eff}^2 = \sum_i \rho_i^2.
\end{equation}
This is what allows for the ``stacking'' of multiple independent events over time: even if the SNR of each individual event is small, by adding up contributions from each event, the \emph{total} log of the Bayes factor can be large, indicating that the evidence for the memory in a population of events is strong.
In this paper, we use $\rho_{\rm eff}^2$ as a proxy for the Bayes factor to make our forecasts less computationally intensive.

\subsection{Memory signals} \label{sec:mem_signal}

The calculation of the Bayes factor above relies on having a procedure to split the model for the waveform into ``oscillatory'' and ``memory'' components.
However, what characterizes the presence of memory (either displacement or spin) in a waveform is the difference in a quantity (respectively the strain or its time integral) before and after the passage of gravitational waves.
For example, the displacement memory is characterized by the difference in the strain $h_{ij}$ between early and late times, and there is no unique way to determine a part of the waveform that contributes to this final difference.
In principle, \emph{any} function which interpolates between the initial and final values of $h_{ij}$ could be a choice for the ``memory'' part of the signal.
However, having a reasonable definition of the memory signal is important, because in detectors such as LIGO, it is the memory part of the signal (however one might define it) that the detectors can measure directly, rather than the finite offset between early and late times.
A similar issue arises for defining the spin memory signal.

However, there is a particularly well-motivated choice that has been used frequently in both numerical relativity (for example,~\cite{Mitman2020}) and the post-Newtonian and post-Minkowskian approximations (see~\cite{Blanchet2013, Nichols2017}).
It relies on the fact that the strain $h_{ij}$ obeys two ``consistency conditions'' that can be derived from the asymptotic form of the Einstein equations in Bondi-Sachs coordinates~\cite{Bondi1962, Sachs1962a} (see, for example,~\cite{Flanagan2015}).
In vacuum, they are given by
\begin{widetext}
\begin{subequations} \label{eqn:conservation}
  \begin{align}
    \ms D^i \ms D^j \Delta h_{ij} (u, u_0) &= \frac{1}{r} \left[4 \Delta m(u, u_0) + \frac{r^2}{2} \int_{u_0}^u \ud u_1 \dot h_{ij} (u_1) \dot h^{ij} (u_1)\right], \label{eqn:cons_m} \\
    \int_{u_0}^u \ud u_1 \ms D^2 \ms D_{[i} \ms D^k h_{j]k} (u_1) &= \frac{1}{r} \bigg\{-4 \ms D_{[i} \Delta \hat N_{j]} (u, u_0) \nonumber \\
    &\hspace{2.8em}+ \frac{r^2}{2} \int_{u_0}^u \ud u_1 \ms D_{[i|} \left[\dot h^{kl} (u_1) \ms D_k h_{l|j]} (u_1) + 3 h_{|j]k} (u_1) \ms D_l \dot h^{kl} (u_1) - (\interchange{h}{\dot h})\right]\bigg\}, \label{eqn:cons_N}
  \end{align}
\end{subequations}
\end{widetext}
where (as is customary) we use square brackets around indices to denote antisymmetrization, with $|$ to indicate indices not antisymmetrized over [so that, for example, $T_{[a|b|c]} \equiv \frac{1}{2} (T_{abc} - T_{cba})$].
Here, $r$ is the distance to the source, the indices $i$, $j$, etc. are indices on the two-sphere, $\ms D_i$ is the covariant derivative on the two-sphere, and $m$ and $N_i$ are functions which appear in the metric called the mass and angular momentum aspects.
Finally, the quantity $\hat N_i$ is defined by
\begin{equation}
  \begin{split}
    \hat N_i (u) &\equiv N_i (u) - (u - u_0) \ms D_i m(u) \\
    &\hspace{1em}- \frac{r^2}{4} \left\{h_{ij} (u) \ms D_k h^{jk} (u) + \frac{1}{4} \ms D_i [h_{jk} (u) h^{jk} (u)]\right\},
  \end{split}
\end{equation}
and $\Delta Q(u, u_0) \equiv Q(u) - Q(u_0)$.
The quantities $m$, $\hat N_i$, and $h_{ij}$ are also functions of the angular coordinates $x^j$, but we suppress the additional functional dependence on $x^j$ to simplify the notation.

For computing the displacement or spin memory effects, the limits of integration in  Eq.~\eqref{eqn:cons_m} or~\eqref{eqn:cons_N} should run from some $u_0$ before the start of the gravitational waves until some $u_1$ after the gravitational waves have passed by, as was the case in, for example,~\cite{Grant2021}.
In that context, the first terms (those involving $\Delta$) on the right-hand side are referred to as the charge contributions, and the second terms are referred to as the (nonlinear) flux contributions, to the displacement or spin memory.

Both equalities in Eq.~\eqref{eqn:conservation} are satisfied for \emph{any} values of $u_0$ and $u$, but in that case they are just a set of expressions that relate the strain to the mass and angular momentum aspects.
When the mass and angular momentum aspects are known as functions of time, Eq.~\eqref{eqn:conservation} can also be used as a consistency check of the waveform (as described in~\cite{Nichols2018, Ashtekar:2019viz}).
If the waveform is found to be inconsistent with Eq.~\eqref{eqn:conservation}, then Eq.~\eqref{eqn:conservation} can be used to determine a correction to the gravitational-wave strain needed to restore consistency.
We use this approach, and the charge-flux-type split on the right-hand sides of Eq.~\eqref{eqn:conservation}, to define the so-called ``memory signals.''

First, for the displacement memory, we define the memory signal to be the part of the strain related to the second term on the right-hand side of Eq.~\eqref{eqn:cons_m}:
\begin{equation} \label{eqn:disp_signal}
  \ms D^i \ms D^j h^{\rm disp.}_{ij} (u) \equiv \frac{r}{2} \int_{u_0}^u \ud u_1 \dot h_{ij} (u_1) \dot h^{ij} (u_1).
\end{equation}
This definition is motivated by the fact that this nonlinear contribution dominates over the one from $\Delta m(u_1, u_0)$ when the spacetime is asymptotically stationary before and after the burst of gravitational waves (that is, outside the interval $[u_0, u_1]$).\footnote{This follows from the fact that the mass aspect $m$ becomes a constant on the sphere and so cannot contribute to $\ms D^i \ms D^j \Delta h_{ij}$, which only contains $l \geq 2$ spherical harmonics~\cite{Flanagan2015}.}
Moreover, it has been confirmed in numerical relativity simulations of binary black holes that the contribution from $\Delta m(u_1, u_0)$ is much smaller than the contribution from Eq.~\eqref{eqn:disp_signal}, at least for the $l = 2, m = 0$ mode~\cite{Mitman2020}.

By analogy, we define the spin memory signal by the integrand of the second term on the right-hand side of Eq.~\eqref{eqn:cons_N}:
\begin{equation} \label{eqn:spin_signal}
  \begin{split}
    \ms D^2 \ms D_{[i} \ms D^k h^{\rm spin}_{j]k} \equiv \frac{r}{2} \ms D_{[i|} \Big[&\dot h^{kl} \ms D_k h_{l|j]} + 3 h_{|j]k} \ms D_l \dot h^{kl} \\
    &- (\interchange{h}{\dot h})\Big].
  \end{split}
\end{equation}
The reason why we consider the integrand in Eq.~\eqref{eqn:cons_N} is that while the displacement memory is related to (differences in the value of) the strain $h_{ij}$, the spin memory is related to its integral.
In contrast, the spin memory \emph{signal} should be the part of the strain that contributes to this integral and hence is given by the integrand.
Similarly to the case of the displacement memory, the contribution to the total integral of the waveform from $\Delta \hat N_i (u_1, u_0)$ has been confirmed to be smaller than the contribution from Eq.~\eqref{eqn:spin_signal} in numerical relativity simulations of binary black holes, at least for the $l = 3, m = 0$ mode~\cite{Mitman2020}.

In principle, the strain $h_{ij}$ and news $\dot h_{ij}$ on the right-hand sides of Eqs.~\eqref{eqn:disp_signal} and~\eqref{eqn:spin_signal} contain $h^{\rm disp}_{ij}$ and $h^{\rm spin}_{ij}$, respectively, thereby making these equations partial integro-differential equations.
However, because it has been determined empirically, for example, in~\cite{Mitman2020} or~\cite{Talbot:2018sgr}, that the oscillatory contribution to $h_{ij}$ is the dominant one, we use the approximation that the terms on the right-hand sides of Eqs.~\eqref{eqn:disp_signal} and~\eqref{eqn:spin_signal} only contain $h^{\rm osc.}_{ij}$.

In this paper, we will often write the strain $h_{ij}$ in terms of spin-weighted spherical harmonics.
Using the plus- and cross-polarization tensors $(e_{+, \times})_{ij}$, we can write \vspace{0.1em}
\begin{equation} \label{eqn:swsh_sum}
  \begin{split}
    h_{ij} [(e_+)^{ij} - i (e_\times)^{ij}] &\equiv h_+ - i h_\times \\
    &\equiv \sum_{\substack{l \geq 2, \\ |m| \leq l}} h_{lm} \, \pb{(-2)} Y_{lm}.
  \end{split}
\end{equation}
The formula for the displacement memory signal in terms of these coefficients of the multipolar expansion of the strain is then
\begin{widetext}
\begin{equation} \label{eqn:disp_modes}
  h_{lm}^{\rm disp.} (u) = \sum_{\substack{l' \geq 2, \\ |m'| \leq l'}} \sum_{\substack{l'' \in \pb{2} I_{ll'mm'}, \\ m'' = m - m'}} \frac{(-1)^{m''} \pb{(-2)2} \mc C^l_{l'l''m'm''}}{\sqrt{(l + 2) (l + 1) l (l - 1)}} \int_{u_0}^u \ud u' \dot h^{\rm osc.}_{l'm'} (u') \overline{\dot h^{\rm osc.}_{l''(-m'')} (u')},
\end{equation}
where
\begin{equation}
  \pb{s} I_{ll'mm'} = \{\max(|s|, |l - l'|, |m - m'|), \cdots, l + l'\},
\end{equation}
and $\pb{ss'} \mc C^l_{l'l''m'm''}$ are coefficients determined by the overlaps of the spin-weighted spherical harmonics:
\begin{equation}
  \pb{s} Y_{lm} \pb{s'} Y_{l'm'} = \sum_{l''} \pb{ss'} \mc C^{l''}_{ll'mm'} \pb{(s + s')} Y_{l''(m + m')}.
\end{equation}
These coefficients can be written in terms of Clebsch-Gordan coefficients (as was done in~\cite{Nichols2017}\footnote{The coefficients $\pb{ss'} \mc C^{l''}_{ll'mm'} $ are equal to the the coefficients $\mathcal C_{l''} (s, l, m; s', l', m')$ in~\cite{Nichols2017}.}), or (equivalently) in terms of Wigner 3-$j$ symbols:
\begin{equation}
  \pb{ss'} \mc C^{l''}_{ll'mm'} = (-1)^{s + s' + m + m'} \sqrt{\frac{(2l + 1) (2l' + 1) (2l'' + 1)}{4\pi}} \begin{pmatrix}
    l & l' & l'' \\
    m & m' & -(m + m')
  \end{pmatrix} \begin{pmatrix}
    l & l' & l'' \\
    -s & -s' & s + s'
  \end{pmatrix}.
\end{equation}
We compute the Wigner 3-$j$ symbols using the software package \texttt{py3nj}~\cite{py3nj}.

The spin memory signal is described in terms of a similar, although somewhat more complicated, equation:
\begin{equation} \label{eqn:spin_modes}
  h^{\rm spin}_{lm} = \sum_{\substack{l' \geq 2, \\ |m'| \leq l'}} \sum_{\substack{l'' \in \pb{2} I_{ll'mm'}, \\ m'' = m - m'}} \frac{\xi^l_{l'l''m'm''} + (-1)^{l + l' + l''} \xi^l_{l''l'm''m'}}{l (l + 1) \sqrt{(l + 2) (l - 1)}} \omega_{l'm'l''m''},
\end{equation}
where\footnote{The coefficients $\xi^l_{l'l''m'm''}$ are related to the coefficients $c^l_{l',m'; l'',m''}$ in~\cite{Nichols2018} by $c^l_{l',m'; l'',m''} = 4 (-1)^{l+l'+l''} \xi^l_{l''l'm''m'}$. }
\begin{equation}
  \xi^l_{l'l''m'm''} \equiv \frac{1}{4} \left[\sqrt{(l' - 2) (l' + 3)} \pb{(-3) 2} \mc C^l{}_{l'l''m'm''} + 3 \sqrt{(l'' + 2) (l'' - 1)} \pb{(-2) 1} \mc C^l{}_{l'l''m'm''}\right]
\end{equation}
and
\begin{equation}
  \omega_{lml'm'} \equiv (-1)^{m'} \left(h^{\rm osc.}_{lm} \overline{\dot h^{\rm osc.}_{l'(-m')}} - \dot h^{\rm osc.}_{lm} \overline{h^{\rm osc.}_{l'(-m')}}\right).
\end{equation}
\end{widetext}
The available waveform models for the oscillatory part of the waveform include only a handful of $lm$ modes, so when we compute the memory effects with these waveforms, the infinite sums will reduce to a sum over a few terms.
We illustrate this with the leading quadrupole approximation for the waveforms in the next part.

\subsubsection{Leading quadrupole-order results}

We now consider an (often-used) approximation, where the only nonzero $h^{\rm osc.}_{lm}$'s are those given by $l = 2$, $m = \pm 2$, and that
\begin{equation} \label{eqn:osc_symm}
  h^{\rm osc.}_{2m} = \overline{h^{\rm osc.}_{2(-m)}} .
\end{equation}
For equal-mass, non-precessing compact binaries, these modes are notably larger than all other multipoles, and thus these modes also contribute the most to the displacement and spin memory signals.
Note that we do \emph{not} use this approximation, and the results of this section, in the main results of this paper.
We only consider this approximation because it shows that certain parts of the oscillatory and memory parts of the waveform are dominant, which is relevant for discussing an issue known as the ``sign of the memory'' in Sec.~\ref{sec:sign}.

Using this approximation, we find that in both Eqs.~\eqref{eqn:disp_modes} and~\eqref{eqn:spin_modes}, $m', m'' = \pm 2$, which requires that $m = 0$ or $m = \pm 4$.
Since $l'' \in \pb{s} I_{ll'mm'}$, it follows that $|l - l'| \leq 2$, and so $2 \leq l \leq 4$.
In order to narrow down the value of $l$, note that, assuming that Eq.~\eqref{eqn:osc_symm} holds,
\begin{equation}
  \dot h^{\rm osc.}_{2m'} \overline{\dot h^{\rm osc.}_{2(-m'')}} = \dot h^{\rm osc.}_{2m'} \dot h^{\rm osc.}_{2m''},
\end{equation}
which is of even parity under $\interchange{m'}{m''}$, while
\begin{equation}
  \omega_{2m'2m''} = (-1)^{m''} \left(h^{\rm osc.}_{2m'} \dot h^{\rm osc.}_{2m''} - \dot h^{\rm osc.}_{2m'} h^{\rm osc.}_{2m''}\right)
\end{equation}
is of odd parity under this transformation, when $m'$ and $m''$ are both even (as they are in this case).
Moreover, the properties of the Wigner 3-$j$ symbols imply that
\begin{equation}
  \label{eqn:C_m_quad_ids}
  \begin{split}
    \pb{ss'} \mc C^l_{22m'm''} &= (-1)^l \pb{ss'} \mc C^l_{22m''m'} \\
    &= (-1)^l \pb{ss'} \mc C^l_{22(-m')(-m'')},
  \end{split}
\end{equation}
which implies that
\begin{equation}
  \label{eqn:xi_m_quad_ids}
  \begin{split}
    \xi^l_{22m'm''} &= (-1)^l \xi^l_{22m''m'} \\
    &= (-1)^l \xi^l_{22(-m')(-m'')}.
  \end{split}
\end{equation}
The first line of Eq.~\eqref{eqn:C_m_quad_ids} implies that the only values of $l$ for which the displacement memory signal does not vanish are $l = 2, 4$ (as found in, for example,~\cite{Wiseman1991}).
The first line of Eq.~\eqref{eqn:xi_m_quad_ids} shows that the only value of $l$ where the spin memory signal does not vanish is $l = 3$~\cite{Nichols2017}.
Similarly, the second lines of Eqs.~\eqref{eqn:C_m_quad_ids} and~\eqref{eqn:xi_m_quad_ids} imply that
\begin{equation} \label{eqn:mem_symm}
  h^{{\rm disp.}/{\rm spin}}_{lm} = (-1)^l \overline{h^{{\rm disp.}/{\rm spin}}_{l(-m)}},
\end{equation}
which (since $\pb{-2} Y_{l0}$ is real) implies that the spin memory signal only has cross polarization, as was found in~\cite{Nichols2017}.

\subsection{``Sign'' of the memory signals} \label{sec:sign}

One of the assumptions made above in Sec.~\ref{sec:stacking} to justify adding the SNRs of the memory signals in quadrature was that the parameters of the binary could be determined precisely from just the oscillatory part of the signal.
However, this will not always be the case, even in the high signal-to-noise limit,  when there are transformations of the parameters of the oscillatory waveform model that leave the waveform invariant (in other words, there are ``degeneracies'' of the waveform model).
More specifically, we are most concerned with cases in which there are different values of the parameters of the model, $\theta$ and $\theta'$, such that $h_{\rm osc.} (\theta)$ and $h_{\rm osc.} (\theta')$ are very close, whereas $h_{\rm mem.} (\theta)$ and $h_{\rm mem.} (\theta')$ are very different.
The case which we consider in this section is where $h_{\rm mem.} (\theta') \approx -h_{\rm mem.} (\theta)$;\footnote{For nonprecessing binaries, the transformation we discuss in Sec.~\ref{subsubsec:transformation} has the property $h_{\rm mem.} (\theta') = -h_{\rm mem.} (\theta)$ as noted in~\cite{Lasky2016,Boersma2020}, but for precessing binaries the relationship between the memory signals at different parameters is approximate, not exact.} namely, there is a parameter degeneracy in the oscillatory waveform model that prevents the sign of the memory signal from being determined.
This degeneracy and its effect on the sign of the memory signal were noted previously in~\cite{Lasky2016}.
We also discuss how this degeneracy can be broken using an approach similar to that in~\cite{Lasky2016,Boersma2020}.

\subsubsection{Transformations of oscillatory and memory signals}
\label{subsubsec:transformation}

We now discuss one such transformation $\theta \to \theta'$ here and discuss a different transformation in Appendix~\ref{app:degeneracy}.
The relevant set of parameters which we consider is
\begin{equation}
  \theta \equiv (\phi_{\rm ref.}, \psi),
\end{equation}
where $\phi_{\rm ref.}$ is the azimuthal coordinate of the detector on the sky (relative to the binary) and $\psi$ (the polarization angle) gives additional information about the orientation of the binary relative to the detector (see, for example,~\cite{Isi:2022mbx}) at a given reference time.
These parameters are discussed in more detail in Sec.~\ref{sec:pops}; see~\cite{LIGO-T1800226-v4} for a diagram showing the definitions of these quantities (where $\phi_{\rm ref.}$ in our notation is denoted by $\phi$, and $\psi$ in our notation is the observable quantity $\psi + \Omega$).
The important relation for this section is the fact that the signal measured by the detector will be
\begin{equation} \label{eqn:swsh_antennas}
  h(\theta) = \sum_{\substack{l \geq 2, \\ |m| \leq l}} h_{(lm)} (\theta),
\end{equation}
where (see~\cite{Boersma2020})
\begin{equation} \label{eqn:antenna_modes}
  \begin{split}
    h_{(lm)} (\theta) &\equiv F_+ (\psi) \Re[h_{lm} \pb{-2} Y_{lm} (\iota, \phi_{\rm ref.})] \\
    &\hspace{1em}- F_\times (\psi) \Im[h_{lm} \pb{-2} Y_{lm} (\iota, \phi_{\rm ref.})],
  \end{split}
\end{equation}
and where $\iota$ is the inclination, or the polar angle of the detector on the sky relative to the binary.
Here, the antenna-pattern functions $F_{+, \times}$ are defined by (see, for example,~\cite{Anderson2000})
\begin{equation}
  F_{+, \times} = D^{ij} (e_{+, \times})_{ij},
\end{equation}
where $D^{ij}$ is a matrix that only depends on the location and orientation of the detector on the earth, and is independent of $\psi$.
All dependence of $F_{+, \times}$ on this parameter arises through $(e_{+, \times})_{ij}$, which are given by
\begin{equation}
  (e_+)_{ij} = X_i X_j - Y_i Y_j, \qquad (e_\times)_{ij} = X_i Y_j + Y_i X_j,
\end{equation}
with
\begin{subequations}
  \begin{align}
    \bs X &= [\sin(\alpha - \omega T) \cos \psi - \cos(\alpha - \omega T) \sin \psi \sin \delta] \bs i \nonumber \\
    &\hspace{1em}+ [-\cos(\alpha - \omega T) \cos \psi - \sin(\alpha - \omega T) \sin \psi \sin \delta] \bs j \nonumber \\
    &\hspace{1em}+ \sin \psi \cos \delta\; \bs k, \\
    \bs Y &= [-\sin(\alpha - \omega T) \sin \psi - \cos(\alpha - \omega T) \cos \psi \sin \delta] \bs i \nonumber \\
    &\hspace{1em}+ [\cos(\alpha - \omega T) \sin \psi - \sin(\alpha - \omega T) \cos \psi \sin \delta] \bs j \nonumber \\
    &\hspace{1em}+ \cos \psi \cos \delta\; \bs k.
  \end{align}
\end{subequations}
Here, $\alpha$ and $\delta$ are the right ascension and declination of the binary on the sky (in equatorial coordinates), $\omega$ is the angular frequency of Earth's rotation, and $T$ is the time of detection; the vectors $\bs i$, $\bs j$, and $\bs k$ are unit vectors in a fixed, inertial, Earth-centered reference frame~\cite{Anderson2000}.

The specific transformation $\theta \to \theta'$ which we consider is given by
\begin{equation} \label{eqn:m_degen}
  \phi_{\rm ref.} \to \phi_{\rm ref.} + \pi/2, \qquad \psi \to \psi + \pi/2.
\end{equation}
Due to the transformation of $\psi$, we find that
\begin{equation}
  \bs X \to \bs Y, \qquad \bs Y \to -\bs X,
\end{equation}
and thus $F_{+, \times} \to -F_{+, \times}$.
Since the spin-weighted harmonics satisfy
\begin{equation}
  \pb{s} Y_{lm} (\iota, \phi_{\rm ref.} + \pi/2) = i^m \pb{s} Y_{lm} (\iota, \phi_{\rm ref.}),
\end{equation}
we find that, for even $m$ (which is the case we are concerned with here)
\begin{equation}
  h_{(lm)} \to -(-1)^{m/2} h_{(lm)}.
\end{equation}
In the case where $h^{\rm osc.}$ is given by only the $l = 2, m = \pm 2$ modes and Eq.~\eqref{eqn:osc_symm} holds, it follows that $h^{\rm osc.}$ is even under this transformation, while $h^{\rm disp.}$ and $h^{\rm spin}$ are both odd.

A key point is that the transformation~\eqref{eqn:m_degen} is only a degeneracy of the oscillatory part of the signal when it is given by the $l = 2$, $m = \pm 2$ modes and obeys Eq.~\eqref{eqn:osc_symm}.
For all compact binaries, there will be additional modes present in the oscillatory signal that are not degenerate; furthermore, if the orbital plane of the binary precesses, Eq.~\eqref{eqn:osc_symm} (and its generalization to higher values of $l$) is also no longer true, which implies that the $l = 2$, $m = \pm 2$ modes are not degenerate under this transformation either.
As we now show, the SNR of the part of the signal that is ``odd'' under such transformations is the part which determines how well $\theta_0$ and $\theta_0'$ can be distinguished.
As we argue in Appendix~\ref{app:sign}, if the parameters cannot be distinguished for a given event, then on average (over different noise realizations) that event contributes much less to the overall Bayes factor $\mc B^{\rm mem.}_{\rm no\,mem.} (d)$.

\subsubsection{Criteria for breaking the degeneracy}
\label{subsubsec:criteria}

To write the criteria for breaking the degeneracy between the parameters, we first write the waveform as
\begin{equation}
  h = h_{\rm even} + h_{\rm odd},
\end{equation}
where
\begin{subequations}
  \begin{align}
    h_{\rm even} (\theta_0') &= h_{\rm even} (\theta_0), \\
    h_{\rm odd} (\theta_0') &= -h_{\rm odd} (\theta_0).
  \end{align}
\end{subequations}
The decomposition of $h$ into even and odd parts applies for the entire waveform (namely, both the oscillatory and memory signals).
Consider now the Bayes factor between the two hypotheses that a set of data $d$ is given by an event with parameters $\theta_0$ and an event with parameters $\theta_0'$:
\begin{equation}
  \begin{split}
    \mc B(d; \theta_0, \theta_0') &= \frac{\exp\left[-\frac{1}{2} \rho_{d - h(\theta_0)}^2\right]}{\exp\left[-\frac{1}{2} \rho_{d - h(\theta_0')}^2\right]} \\
    &= \exp\left[2 \<d - h_{\rm even} (\theta_0)|h_{\rm odd} (\theta_0)\>\right].
  \end{split}
\end{equation}
We assume here that the prior probabilities satisfy $p(\theta_0) = p(\theta'_0)$.
Note that, unlike in Eq.~\eqref{eqn:bayes_factor}, there is not an integral over the different parameters, as the particular values of the parameters are part of the hypothesis.
Using the fact that $d = h(\theta_0) + n$, we find [upon averaging over the noise using Eq.~\eqref{eqn:avg_exp}] that
\begin{equation}
  \ln \expect[\mc B(d; \theta_0, \theta_0')] = 4 \rho_{h_{\rm odd} (\theta_0)}^2.
\end{equation}
This shows that the SNR of the odd part of the signal is what determines how well this degeneracy can be broken.
Moreover, since the Bayes factor for independent measurements is multiplicative, this means that the effective SNR$^2$ is additive, just as it was in the case of the detection of the memory.
This justifies adding together the SNR$^2$ for the measurements of the same event by multiple detectors, which makes this degeneracy easier to break.
In this paper, we assume that an SNR$^2$ of 2 in the odd part of the oscillatory signal, computed for the \emph{entire} detector network, is sufficient to break the degeneracy.\footnote{In~\cite{Lasky2016}, the definition of $h_{\rm odd}$ was twice that of our definition, so their threshold of 2 for the SNR would be equivalent to a threshold of 1 by our definition of $h_{\rm odd}$.
Note, however, that their $h_{\rm odd}$ contained just $l = 2$ and $l = 3$ modes, whereas ours includes all the available $(l, m)$ modes in the waveform models that we use.
  In~\cite{Boersma2020}, only the odd $m$ modes were used in the definition of the higher-order-mode waveform $h_{\rm HOM}$, which was used to compute an SNR threshold.
  A short calculation shows that $\rho_{h_{\rm odd}}^2 \simeq \rho_{h_{\rm HOM}}^2/2$; thus, the threshold chosen in~\cite{Boersma2020} of $\rho_{h_{\rm HOM}} = 2$ is approximately equivalent to the choice that we have made of $\rho_{h_{\rm odd}}^2 = 2$.}
Using $h_{\rm osc.}^{\rm odd}$ rather than the full $h_{\rm odd}$ to break the degeneracy does not make much difference, because the SNR of the memory component $h_{\rm mem.}^{\rm odd}$ is typically a small correction relative to the SNR of the oscillatory component, and the memory and oscillatory waveforms have a negligible overlap (noise-weighted inner product).

Finally, we note that having a large value for the SNR for the odd part of the oscillatory signal does not just break this degeneracy.
The result that the displacement and spin memory signals are entirely odd under the transformation in Eq.~\eqref{eqn:m_degen} relies on the oscillatory part of the signal entirely being in the $l = 2, m = \pm 2$ modes and the fact that Eq.~\eqref{eqn:osc_symm} holds.
If there were a case in which the odd part of the oscillatory signal is large, then the displacement and spin memory signals could also have a large even part compared with their odd parts.
We did not note any binaries in our simulated populations that had this property, however.

\section{Forecasting methods} \label{sec:methods}

We outline several different aspects of our forecasting procedure in this section.
We begin by describing the population models and how we draw realizations of these populations from these models in Sec.~\ref{sec:pops}.
We then discuss in Sec.~\ref{sec:StoN} the signal-to-noise ratio calculations, in particular the factors that influence it, such as the detector networks, the waveform families, and the signal processing methods used.

\subsection{Populations and events} \label{sec:pops}

To forecast the prospects for detecting gravitational wave memory effects, we first must generate simulated populations of events that are consistent with the binary black hole mergers that have been detected so far.
Each event is characterized by a set of 15 parameters.
The intrinsic parameters (that is, those which do not depend on the location of the detector) are
\begin{enumerate}

\item the mass $m_1$ of the primary;

\item the mass ratio $q \equiv m_2/m_1$;

\item the (dimensionless) spin magnitudes $\chi_1$ and $\chi_2$;

\item $z_1$ and $z_2$, the cosines of the angle of each spin vector relative to the orbital plane; and

\item $\phi_1$ and $\phi_2$, the angles of each spin vector in the orbital plane relative to some arbitrarily chosen axis.

\end{enumerate}
Because the spins and the orbital plane precess, $z_{1, 2}$ and $\phi_{1, 2}$ need to be specified at some time.
By convention, this is when the $l = 2, m = \pm 2$ modes (in the co-precessing frame) have a ``reference frequency'' $f_{\rm ref.}$, which is often related to the low-frequency cut-off of the detector.
The remaining extrinsic parameters are
\begin{enumerate}

\item the location of the detector on the sky, relative to the plane of the orbit, determined by the polar angle $\iota$ and axial angle $\phi_{\rm ref.}$;

\item the location of the binary on the sky, relative to the detector, as determined by the right-ascension $\alpha$ and the declination $\delta$;

\item the polarization angle $\psi$, determining the orientation of the binary on the sky relative to a fixed polarization frame defined on Earth;

\item the redshift\footnote{More precisely, it is the luminosity distance $D_L$ and redshifted masses that appear in the amplitude of the gravitational waveform; the redshift is then inferred from the luminosity distance assuming a given cosmological model. However, Ref.~\cite{LSC2020c} parametrized the distances in terms of the redshift $z$.} $z$; and

\item the time difference $\Delta T$ between each event, which can be summed to determine the absolute (GPS) time $T$ that is frequently used by ground-based detectors.

\end{enumerate}
Note, again, that the orbital plane used to define $\iota$ and $\phi_{\rm ref.}$ is the orbital plane at $f_{\rm ref.}$.
Moreover, $\phi_{\rm ref.}$ is the axial angle that is relative to the vector pointing towards the primary black hole from the secondary, again at $f_{\rm ref.}$.

\subsubsection{Population model} \label{sec:pop_model}

To generate simulated populations of events, we need a model for the distributions of these parameters.
Several parameters, by assumptions of isotropy, will have either uniform distributions ($\phi_1$, $\phi_2$, $\phi_{\rm ref.}$, $\alpha$, and $\psi$), or effectively uniform distributions ($\cos \iota$ and $\sin \delta$ are uniformly distributed).
The distributions of the remaining event parameters are determined by the particular population model in question.
The population model that we use to generate events in this paper is the so-called ``Power Law + Peak'' (PLPP) model~\cite{Talbot2018a}, which is the model that was most favored by the events during the O1, O2, and O3a observing runs of LIGO and Virgo, which form the second gravitational wave transient catalog (GWTC-2)~\cite{LSC2020c}.
This population model is characterized by a set of population parameters, and the observed events were used to determine posterior distributions for these parameters using a hierarchical Bayesian analysis~\cite{LSC2020c}.

We now consider the various pieces of this model.
The PLPP model describes the distribution of the primary mass and mass-ratio of the merging binary black holes, and it is characterized by 9 population parameters:\footnote{Reference~\cite{LSC2020c} described the PLPP model as depending on only 8 parameters, because one of the 9 parameters was fixed to a constant value and not inferred from the observed black hole mergers; see Footnote~\ref{fn:gauss_max}.}
\begin{enumerate}

\item $m_{\rm min}$, which is the minimum value of the mass of either black hole in the binary;

\item $m_{\rm max}$, which is the maximum mass of either black hole for the power-law component of the mass distribution;

\item $\delta_m$, which characterizes the width of the transition of the mass distribution at low masses to zero;

\item $\alpha$ and $\beta$, which are power law exponents characterizing the power law component of the primary mass distribution and the mass-ratio distribution, respectively;

\item $\mu_m$ and $\sigma_m$, which give the mean and width of the Gaussian component of the primary mass distribution;

\item $m_{\rm gauss. max}$, which determines the upper mass of the Gaussian component, and also determines the Gaussian's normalization;\footnote{Note that we did not see this value specified explicitly in either of the references~\cite{Talbot2018a, LSC2020c}.
    The source code of the \texttt{gwpopulation} package~\cite{Talbot2019} (mentioned as being used by~\cite{LSC2020c}) indicates that the Gaussian component is normalized over $[m_{\rm min}, 100 M_\odot]$.
    In a private communication, Colm Talbot confirmed this choice of normalization.
    We therefore fix $m_{\rm gauss. max} = 100 M_\odot$ in our analysis.
    \label{fn:gauss_max}} and

\item $\lambda_{\rm peak}$, which determines fraction of systems in the Gaussian component of the primary mass distribution.

\end{enumerate}
The form of the joint $m_1$-$q$ probability distribution is
\begin{widetext}
\begin{equation}
  \begin{split}
    \pi(m_1, q) &\propto P(q; \beta, m_{\rm min}/m_1, 1) \left[(1 - \lambda_{\rm peak}) P(m_1; -\alpha, m_{\rm min}, m_{\rm max}) + \lambda_{\rm peak} G(m_1; \mu_m, \sigma_m, m_{\rm min}, m_{\rm gauss. max})\right] \\
    &\hspace{1em}\times S(m_1; m_{\rm min}, \delta_m) S(q m_1; m_{\rm min}, \delta_m),
  \end{split}
\end{equation}
where
\begin{equation}
  P(x; \gamma, x_0, x_1) \equiv \frac{\gamma + 1}{x_1^{\gamma + 1} - x_0^{\gamma + 1}} \begin{cases}
    x^\gamma & x_0 \leq x \leq x_1 \\
    0 & \textrm{otherwise}
  \end{cases}
\end{equation}
is a power law distribution with exponent $\gamma$, normalized on the domain $[x_0, x_1]$;
\begin{equation}
  G(x; \mu, \sigma, x_0, x_1) \equiv \frac{1}{\sigma \sqrt{\frac{\pi}{2}} \left[\erf \left(\frac{x_1 - \mu}{\sigma \sqrt{2}}\right) - \erf \left(\frac{x_0 - \mu}{\sigma \sqrt{2}}\right)\right]} \begin{dcases}
    \exp\left[-\frac{(x - \mu)^2}{2 \sigma^2}\right] & x_0 \leq x \leq x_1 \\
    0 & \textrm{otherwise}
  \end{dcases}
\end{equation}
is a truncated Gaussian normalized on the domain $[x_0, x_1]$ \{and $\erf(y)$ is the error function; see, for example, Eq.~(7.2.1) of~\cite{NIST:DLMF}\}; and
\begin{equation} \label{eqn:smooth}
  S(x; x_0, \delta) \equiv \begin{dcases}
    0 & x \leq x_0 \\
    \left\{1 + \exp\left[\frac{\delta}{x - (x_0 + \delta)} + \frac{\delta}{x - x_0}\right]\right\}^{-1} & x_0 \leq x \leq x_0 + \delta \\
    1 & x \geq x_0 + \delta
  \end{dcases}
\end{equation}
\end{widetext}
is a smooth transition function between $0$ and $1$ from $x_0$ to $x_0 + \delta$.

Next, we consider distributions for the parameters related to the spin (ignoring $\phi_1$ and $\phi_2$, which are sampled from a uniform distribution).
We use the so-called ``default'' spin model in~\cite{LSC2020c}, in which the spin magnitudes $\chi_1$ and $\chi_2$ are drawn from beta distributions and the parameters $z_1$ and $z_2$ are drawn from a sum of a uniform distribution and a product of Gaussians~\cite{Wysocki2018, Talbot2017}.
The population parameters are
\begin{enumerate}

\item $\mu_\chi$ and $\sigma_\chi$, the mean and standard deviation of the spin magnitude distributions, which are the same for both spins;

\item $\sigma_t$, the standard deviation of the Gaussian component of the $z_1$ and $z_2$ distribution; and

\item $\zeta$, the fraction of the distribution for $z_1$ and $z_2$ that is in the Gaussian component.

\end{enumerate}
Explicitly,
\begin{equation}
  \pi(\chi_i) \propto \chi_i^{\alpha_\chi - 1} (1 - \chi_i)^{\beta_\chi - 1},
\end{equation}
where $\alpha_\chi$ and $\beta_\chi$ are defined by
\begin{equation}
  \alpha_\chi \equiv \frac{(1 - \mu_\chi)}{\sigma_\chi^2} - \frac{1}{\mu_\chi}, \qquad \beta_\chi \equiv \alpha_\chi \left(\frac{1}{\mu_\chi} - 1\right).
\end{equation}
The distribution for $z_1$ and $z_2$ is given by
\begin{equation}
  \begin{split}
    \pi_{z_{1, 2}} (z_1, z_2) &\propto \zeta G(z_1; 1, \sigma_t, -1, 1) G(z_2; 1, \sigma_t, -1, 1) \\
    &\hspace{1em}+ \frac{1 - \zeta}{4}.
  \end{split}
\end{equation}

Finally, we use a model for redshift and time $\Delta T$ between events where the event rate evolves with redshift~\cite{Fishbach2018}.
This model is characterized by the following parameters:
\begin{enumerate}

\item $R_0$, the local ($z = 0$) event rate (with respect to detector time), per comoving volume;

\item $\kappa$, the power law exponent for the event rate per comoving volume; and

\item $z_{\rm max}$, the maximum value of redshift for binary black hole mergers allowed for in this model.

\end{enumerate}
Unless otherwise specified, we take $z_{\rm max} = 1$, and do not draw any events at higher redshifts.
While astrophysical events will occur at higher redshifts, this particular population model is only well-constrained up to $z = 1$~\cite{LSC2020c}.
Note that systems may be detected at greater values of redshift during the O4 and O5 observing runs, and will be detected at much larger redshifts by Cosmic Explorer~\cite{Evans2021}.
However, since the star formation rate peaks at $z \simeq 1.9$~\cite{Madau2014}, simply extending these models may not correctly estimate the number of events at redshifts higher than one.
In an attempt to be conservative in our forecasts, we do not extrapolate these models and simply use $z_{\rm max} = 1$.

The explicit distribution for redshift is given by a power law
\begin{equation}
  \pi(z) \propto \begin{dcases}
    (1 + z)^{\kappa - 1} \frac{\ud V_c}{\ud z} & 0 \leq z \leq z_{\rm max} \\
    0 & \textrm{otherwise}
  \end{dcases},
\end{equation}
where $\ud V_c/\ud z$ is the differential comoving volume.
The distribution for $\Delta T$ is that given by a Poisson process:
\begin{equation}
  \pi(\Delta T) \propto \exp\left[-\frac{\Delta T}{R_0 \int_0^{z_{\rm max}} (1 + z)^{\kappa - 1} \frac{\ud V_c}{\ud z} \ud z}\right].
\end{equation}
Note that $\ud V_c/\ud z$, as a function of redshift, depends on the cosmological model which we are using; here, we use the software package \texttt{astropy}~\cite{Astropy2013, Astropy2018}, using their \texttt{Planck15} model based on Planck 2015 cosmological parameters~\cite{Planck2015}.

\subsubsection{Drawing population parameters and events}

With the population model now defined, we turn to discussing how we draw the population parameters from the posterior distributions given in the public data release of~\cite{LSC2020c}, available at~\cite{LIGO-P2000434}.
These distributions are given as a set of samples which represent a fair draw from the full posterior distribution.
We do not want to be restricted to this specific draw from the distribution, so we instead use the software package \texttt{kombine} to draw from an estimate of the full distribution that was constructed from a Gaussian kernel density estimation~\cite{kombine}.
Once we have generated a set of parameters for a population, we draw event parameters in order to generate a set of events.
For $m_1$, $q$, and $z$, there are no simple analytic methods to draw from their distributions, so we use Markov chain Monte Carlo (MCMC) sampling as implemented in the \texttt{emcee} software package~\cite{emcee2012}.
Note that there are only two separate distributions that need to be sampled: the joint distribution for $m_1$ and $q$ and the distribution for $z$.
The remaining parameters can all be drawn ``exactly,'' as there exist efficient methods to draw from their distributions using random number generators.~\footnote{That the joint distribution for $z_1$ and $z_2$ can be sampled exactly follows from the fact that it is the sum of two distributions, each of which can also be sampled exactly.
  That this is the case was pointed out to us by an anonymous referee.
  \label{fn:sampling}}

Finally, the parameters which characterize events in the population model described above are not those which most naturally arise when generating waveforms.
We must transform these parameters into the following parameters, which can be computed as functions of the event parameters given above:
\begin{enumerate}

\item the redshifted total mass $M \equiv (1 + q) (1 + z) m_1$;

\item the luminosity distance $D_L (z)$;\footnote{This function is dependent on the cosmological model in question; as we did with $\ud V_c/\ud z$ above, we use the \texttt{Planck15} model of \texttt{astropy}~\cite{Astropy2013, Astropy2018}.}

\item the Cartesian spin vectors $\vec{\chi}_{1, 2}$, which are constructed from the magnitudes and angles using the transformations from spherical polar to Cartesian coordinates; and

\item the GPS time $T$ of the event relative to some reference time, determined by adding up the $\Delta T$ of each previous event.

\end{enumerate}
These ``waveform parameters'' allow us to compute the signal-to-noise ratio of each event.

\subsection{Signal-to-noise ratio of single events} \label{sec:StoN}

In this section, we describe the computation of the SNR of each event.
As described above, for each event we compute the SNR of the ``odd'' part of the waveform under the transformation in Sec.~\ref{sec:sign} (to determine the sign of the memory), and then we use that as a threshold for whether or not to compute the SNR of the displacement or spin memory signal.
Events for which the SNR$^2$ of the odd part of the waveform exceeds the threshold of two also have a sufficiently high SNR that they would pass the threshold of detection for the total signal.
We therefore do not impose a minimum SNR for the total signal.

\subsubsection{Detector networks}

Here, we describe the various parameters which we use in order to model the (network of) detectors used to measure these signals.
We summarize these parameters in Table~\ref{tab:networks}.

\begin{table*}[htb]
  \caption{\label{tab:networks} A summary of the detector networks considered in this paper.
    ``HLVK'' refers to the second-generation detector network of LIGO (Hanford and Livingston), Virgo, and KAGRA; ``HLVKI''  denotes HLVK plus LIGO India.
    ``2CE'' and ``3CE'', respectively, refer to networks with two and three Cosmic Explorer detectors, where 2CE has detectors in the United States and Australia, and 3CE has an additional detector in Europe.
    We also give the references that we use to compute the amplitude spectral densities $\sqrt{S_n (f)}$ and the antenna response functions $F_{+/\times}$.
    See the main text for a justification for the run time for each detector network.}
  \bgroup
  \setlength{\tabcolsep}{1em}
  \begin{tabular}{lllll}
    Detector network & Frequency range (Hz) & $\sqrt{S_n (f)}$ & $F_{+, \times}$ & Run time (yrs) \\\hline\hline
    HLVK O4 & $[10, 4096]$ & \cite{LIGO-T2000012-v2}\footnote{Uses the ``high'' sensitivity for KAGRA in O4} & \cite{lalsuite} & 1.5 \\
    HLVK O5 & \ditto & \ditto\footnote{Uses the ``high'' sensitivity for Virgo in O5 \label{fn:virgo_O5}} & \ditto & 0.75 \\
    HLVKI O5 & \ditto & \ditto\footnote{LIGO India is considered here to be a copy of LIGO Livingston or Hanford} & \ditto\footnote{Location for LIGO India in~\cite{lalsuite} is only a hypothetical example} & 2.75 \\\hline
    2CE/3CE & $[5, 4096]$ & \cite{CE-T2000017-v5}\footnote{Sensitivity is for the ``default'' mode and an arm length of 40 km, comparable to the Stage 2 sensitivity of~\cite{Reitze2019}} & \cite{Hall2019} & 5 \\\hline\hline
  \end{tabular}
  \egroup
\end{table*}

As described above Eq.~\eqref{eqn:swsh_antennas}, the signal measured by a single detector depends on the antenna pattern functions $F_{+}$ and $F_{\times}$.
In the frequency domain, we have that
\begin{equation}
  \tilde h(f) = F_+ \tilde h_+ (f) + F_\times \tilde h_\times (f),
\end{equation}
where we have assumed that the time-dependence in the antenna pattern functions can be ignored; this approximation is only valid for sufficiently short signals (significantly shorter than a day).
With the frequency-domain signal, we estimate the SNR (squared) by its optimal value, which is given by the integral:
\begin{equation} \label{eqn:snr_approx}
  \rho_h^2 = 4 \int_{f_{\rm low}}^{f_{\rm high}} \ud f \frac{|\tilde h (f)|^2}{S_n (f)},
\end{equation}
integrating only between $f_{\rm low}$ and $f_{\rm high}$.

For a given event, multiple SNRs from a collection of detectors can be added in quadrature, as described in Sec.~\ref{sec:stacking}.
Each individual detector has a ``duty cycle,'' which reflects the fraction of time that the detector is operational.
As a pessimistic estimate based upon O3, we pick a duty cycle for O4 and O5 of 75\%~\cite{Buikema2020}, which we also use for Cosmic Explorer.
Moreover, for each event, we consider the total SNR to be zero if only one detector is operating when it occurs.
This prescription is how we compute the total SNR for an event in a particular network of detectors.

Detector networks improve in sensitivity over time, and by using different networks, we can show how improvements in detector sensitivity will affect the detection prospects for the different memory effects.
The detector upgrade timescale is best estimated for second-generation detectors, and the projected detector sensitivity will have an important impact on the memory detection forecasts.
The particular choice of run times for O4 and O5 (with and without LIGO India) are given by the observation scenarios in~\cite{KAGRA2013}.
Note, however, that~\cite{KAGRA2013} did not give observing scenarios after two years of operation of O5.
We nevertheless run the full, final network (O5 with LIGO India) for a longer period of time than this so that our forecasts extend for 5 years of O4 and O5 operations.
This duration was chosen to allow for a direct comparison with the results~\cite{Boersma2020} (and it will prove long enough to have a good chance of detecting the displacement memory effect).
For Cosmic Explorer, we also use 5 years, which is a sufficiently long period of time to potentially detect the spin memory effect, and it is on the order of the expected run time of Cosmic Explorer~\cite{Reitze2019}.

\subsubsection{Waveform generation}

We now discuss the waveform families and approximations that we use for the oscillatory and memory signals that enter into the signal-to-noise ratio calculations.
Since the population described in Sec.~\ref{sec:pops} includes spins, the model used to generate these waveforms needs to be able to account for nonzero spins and, in general, spins that are not aligned with the orbital angular momentum.
Such spins result in a precession of the orbital plane; this results in a significant contribution to the ``odd'' part of the waveform, the SNR of which is used as a threshold for determining the sign of the memory effect.

In this paper, we make use of three different oscillatory waveform models:
\begin{enumerate}

\item an effective one-body (EOB) model \texttt{SEOBNRv4PHM}~\cite{Ossokine2020}, which is available as part of the \texttt{lalsimulation} software package~\cite{lalsuite};

\item a numerical relativity surrogate model \texttt{NRSur7dq4}~\cite{Varma2019}, available through the \texttt{gwsurrogate} software package~\cite{Field2013}; and

\item a hybridized numerical relativity surrogate model \texttt{NRHybSur3dq8}~\cite{Varma2018}, also available through \texttt{gwsurrogate}.

\end{enumerate}
Each of these models has advantageous and disadvantageous features, which we summarize in Table~\ref{tab:waveforms}.

\begin{table*}[htb]
  \caption{\label{tab:waveforms} A summary of the waveform models considered in this paper.
    In particular, it shows that there are trade offs between allowing for arbitrary spins, including the early inspiral, and being able to be evaluated as rapidly as possible.
    After a comparison described in more detail in the text, we settled upon using \texttt{NRSur7dq4} for all calculations with mass ratios with $q > 1/4$ and \texttt{SEOBNRv4PHM} otherwise.}
  \bgroup
  \setlength{\tabcolsep}{1em}
  \begin{tabular}{lccc}
    Waveform model & Arbitrary spins? & Early inspiral? & Run time (s)\footnote{Computed for an event with the same parameters as in Footnote~\ref{fn:params}, with a total mass $M = 65 M_\odot$ and a starting frequency of $10$ Hz (for \texttt{SEOBNRv4PHM} and \texttt{NRHybSur3dq8})} \\\hline\hline
    \texttt{SEOBNRv4PHM} & Yes & Yes & $\sim 4.2$ \\
    \texttt{NRSur7dq4} & Yes & No & $\sim 0.091$ \\
    \texttt{NRHybSur3dq8} & No & Yes & $\sim 0.26$ \\\hline\hline
  \end{tabular}
  \egroup
\end{table*}

\begin{figure}
  \includegraphics[width=\linewidth]{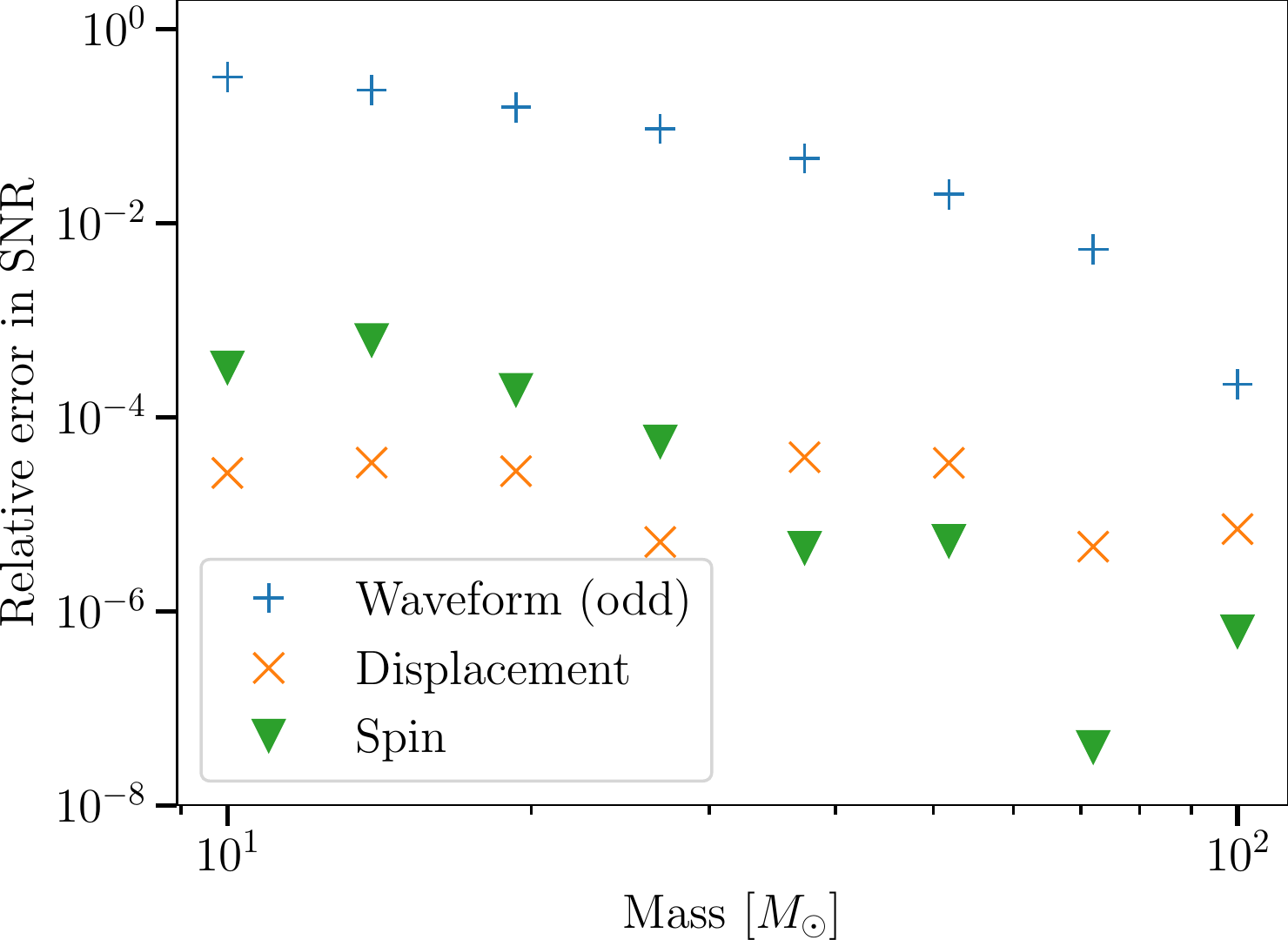}
  \caption{\label{fig:length} A plot comparing the relative error in the SNR computed either from the full waveform or a waveform truncated in length.
    The specific SNRs that are computed are from the parts of the waveform that are odd under the transformation in Sec.~\ref{sec:sign}, the displacement memory signal, and the spin memory signal.}
\end{figure}

For the analysis of this paper, except outside of the range of mass-ratios in which it is valid, we use the (non-hybridized) surrogate model \texttt{NRSur7dq4}; for $q < 1/4$, we use \texttt{SEOBNRv4PHM}.
This is motivated first by considering a collection of zero-spin events where we fix the values of all the parameters except for the mass $M$ of the primary.\footnote{The particular values which we use are $q = 0.8$, $\iota = 7\pi/9$, $D_L = 410$ Mpc, with all of the remaining parameters (except for $M$, which varies) set to zero.\label{fn:params}}
The non-hybridized surrogate model \texttt{NRSur7dq4} will have some fixed length which, for masses smaller than $\sim 100 M_\odot$, is shorter than the hybridized surrogate model \texttt{NRHybSur3dq8}.
In Fig.~\ref{fig:length}, we compare the SNRs (as computed in the HLVKI network of detectors at its O5 sensitivity) of the odd part of the waveform, the displacement memory signal, and the spin memory signal, computed first using the full waveform from \texttt{NRHybSur3dq8}, and then using a truncated waveform of the same length as the \texttt{NRSur7dq4} waveform.
This figure shows that for the masses considered, the length of the waveform affects the SNR of the displacement memory signal by at most a hundredth of a percent, and that of the spin memory signal by a tenth of a percent.
Note that this result is somewhat surprising, as the amplitude of the final displacement memory \emph{does} depend strongly on the length of the waveform; however, the SNR does not.
The SNR of the odd part of the waveform depends much more strongly on the length of the waveform; however, as computing the SNR of this part of the waveform is performed for \emph{every} event, using \texttt{SEOBNRv4PHM} would be prohibitively expensive.
As such, we use \texttt{NRSur7dq4} for all calculations in this paper, only using \texttt{SEOBNRv4PHM} when $q < 1/4$.
From Fig.~\ref{fig:length} we anticipate that using different waveform models will affect the results of these forecasts by at most a few tens of percent, but a thorough investigation of these effects is outside the scope of the current work.

Finally, we choose a sampling frequency for these waveforms of $f_{\rm samp.} = 2 f_{\rm high}$, which in all cases is $8192$ Hz.
Moreover, for the ``reference frequency'' remarked upon above, we use $f_{\rm ref.} = 0.03/M$.
Note that this is not the same as the starting frequency for any of the waveforms which we use: as remarked above, it is used to determine the time at which the orbital plane, the components of the spins, etc., are defined.
We choose a specific value of $f_{\rm ref.}$, instead of defining these quantities at the starting frequency, since the starting frequencies vary.\footnote{Note that the analysis used to generate the distribution of parameters for the populations in~\cite{LSC2020c} may not have used such a uniform value of this reference frequency, but we suspect that this difference should not significantly affect our results.}

\subsubsection{Windowing of waveform time series}

The waveforms discussed above are all generated in the time domain, whereas the SNR is computed in the frequency domain.
It is therefore necessary to compute the Fourier transform of all of the time-domain signals.
The Fourier transform is estimated using an implementation (in \texttt{NumPy}~\cite{NumPy}) of the fast Fourier Transform (FFT) algorithm.
Since the FFT estimates from a finite-length time series the Fourier transform of a formally infinite-length time series, it is necessary to ensure that any finite-time effects do not influence the estimate of the Fourier transform in the frequency bandwidth of the detector.
Some methods for mitigating these finite-time effects are discussed in~\cite{LSC2019b}.
We will briefly describe a few issues here that are most pertinent for our memory forecasts.

Since the signals which we consider are nonzero at the boundaries of the domain (either the start of the waveform, for the odd part of the waveform, or at the end of the waveform for the displacement memory signal), the periodicity inherent in the FFT treats this nonzero value as a discontinuity and leads to ``edge effects.''
The known way to diminish the effects of this discontinuity is to apply a time-domain window function that smooths out this discontinuity (and thus avoids spurious $1/f$ features in the Fourier transform).

\begin{figure}
  \includegraphics[width=\linewidth]{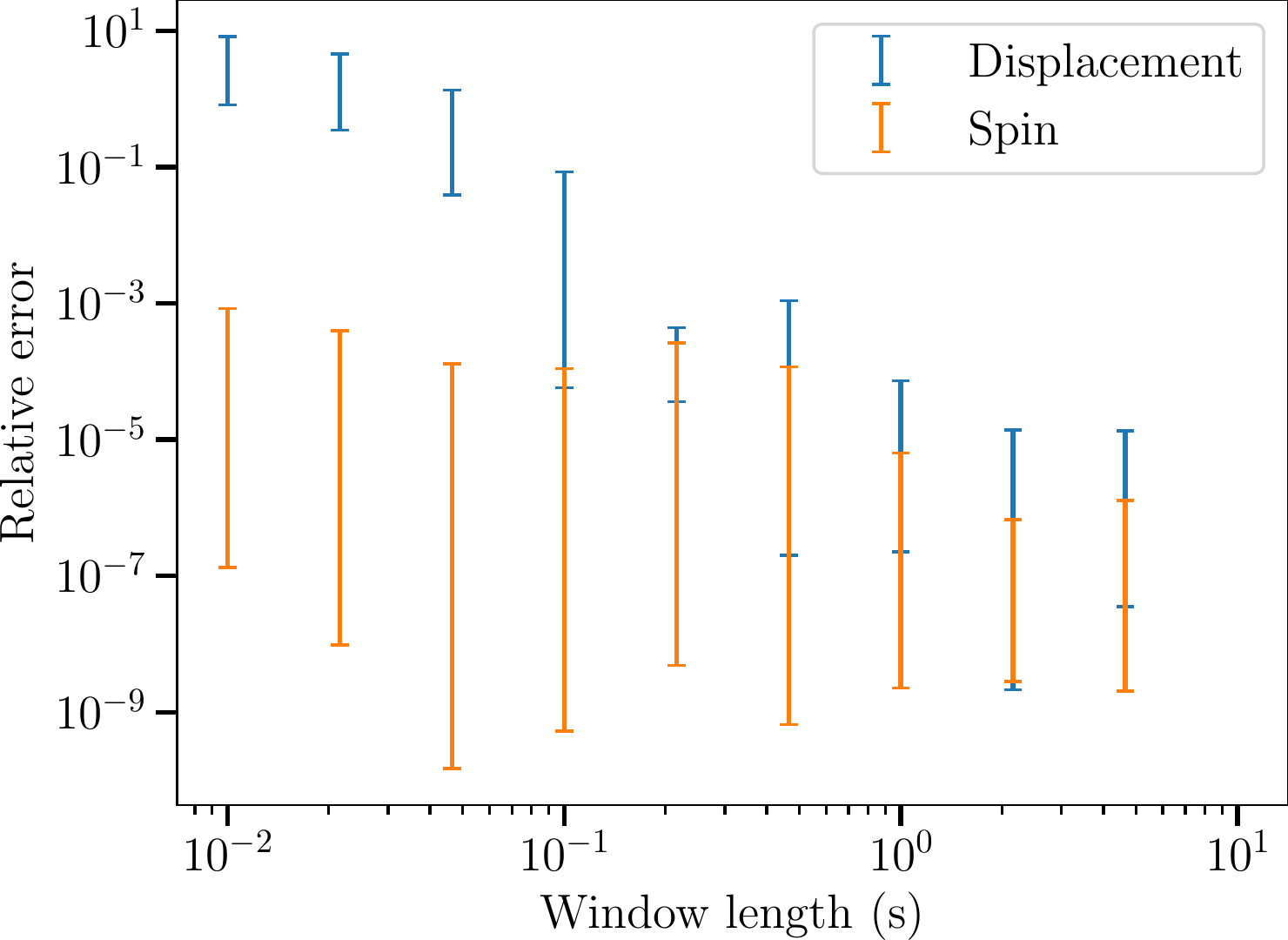}
  \caption{\label{fig:mem_window} The relative error in the SNR vs. window length, for just the displacement and spin memory signals, using a window that pads both the beginning and end of the waveform.
    The error is computed by comparing against a window with a length of $10$s.
    The SNR is computed in the HLVKI detector at its O5 sensitivity, and for a range of masses between $10$ and $10^{2.5} M_\odot$; the remaining parameters are the same as in Footnote~\ref{fn:params}.}
\end{figure}

\begin{figure}
  \includegraphics[width=\linewidth]{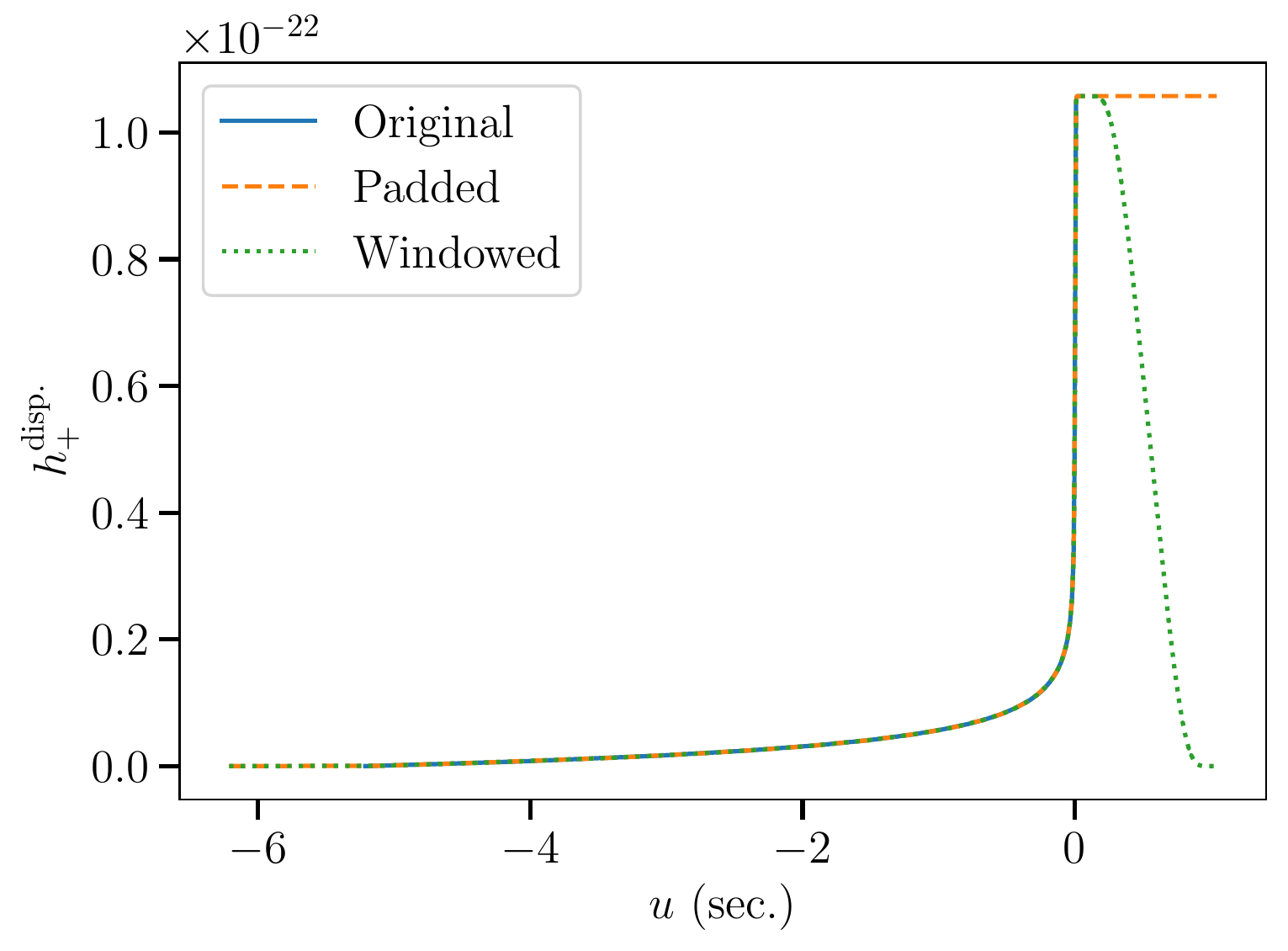}
  \caption{\label{fig:window_example}
    The displacement memory signal, as a function of time, showing the method of padding and windowing described in the text.
    For this particular case, we only show the $+$-polarized component of the displacement memory signal at the location of the detector.
    The mass of the system is $65 M_\odot$, and the remaining parameters are the same as in Footnote~\ref{fn:params}.}
\end{figure}

In Fig.~\ref{fig:mem_window}, we show the effects of windowing on the SNR of the displacement and spin memory signals.
This windowing is performed by first padding both the beginning and end of the waveform (using the values at the beginning or end, in order to preserve continuity), and then applying a window over both of the padded portions of the waveform.
An explicit example of the windowing applied to the memory signal is given in Fig.~\ref{fig:window_example}.
Throughout this paper, we use the Planck window~\cite{McKechan2010}, which uses the same smoothing function as in Eq.~\eqref{eqn:smooth} at the start of the window (and a similar function at the end of the window).
As one would expect that an infinitely long window is closest to the true Fourier transform of the signal, we compute in Fig.~\ref{fig:mem_window} the error in the SNR, relative to a window of length $10$s.
This figure shows that padding and windowing over $1$s is sufficient for the range of masses which will be important for this paper, and as that is less computationally expensive, we use this window length for the results of this paper.
It also demonstrates that not windowing the displacement memory signal results in a significant relative error.

\begin{figure}
  \includegraphics[width=\linewidth]{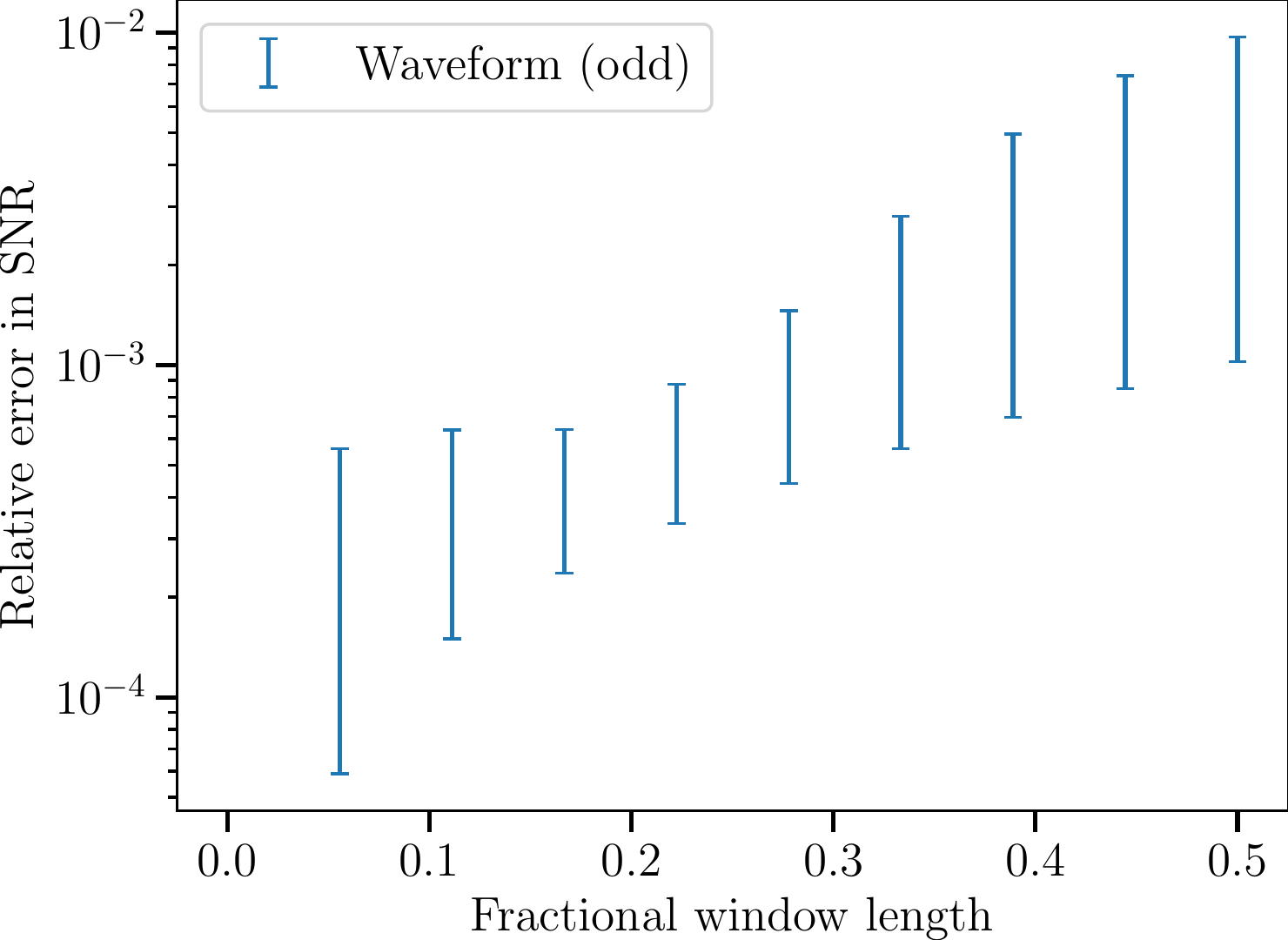}
  \caption{\label{fig:wf_window} The relative error in the SNR, as a function of the length of the window (as a fraction of the length of the unwindowed waveform), for the odd part of the waveform.
    The window is applied over both the beginning of the waveform and a padded region at the end of the waveform of the same length.
    The error is computed by comparing with a window length of $0$.
    As in Fig.~\ref{fig:mem_window}, the SNR is computed in the HLVKI detector at its O5 sensitivity, for a range of masses between $10$ and $10^{2.5} M_\odot$; the parameters are the same as in Footnote~\ref{fn:params}.}
\end{figure}

We now turn to the windowing of the odd part of the waveform, the effects of which are shown in Fig.~\ref{fig:wf_window}.
Here, since padding the beginning of the signal would result in a signal with a large discontinuity in its first derivative, we instead only pad the end of the waveform by some fraction of the waveform's length.
We then apply a window with the same length as this padded portion of the waveform, so that the beginning of the waveform is also windowed over.
This plot computes the error in the SNR as a function of the fraction of the waveform's length over which the window is applied, relative to a window of length zero (so no windowing at all).
Since the effects of windowing are very small, in this paper we do \emph{not} window the odd part of the waveform when computing its SNR to speed up our computation.

\section{Forecast results} \label{sec:results}

In this section, we present the main results of this paper, which are contained in Figs.~\ref{fig:2g_disp} and~\ref{fig:3g_spin}.
Figure~\ref{fig:2g_disp} shows the accumulated SNR in second-generation HLVK and HLVKI detector networks for the displacement memory signals; Fig.~\ref{fig:3g_spin} shows the signal-to-noise ratio for the spin memory signal in two and three Cosmic Explorer detectors at their full sensitivities.
To generate these forecasts, 300 realizations of populations of events are used, as in~\cite{Boersma2020}.
We then compute the median values and credible regions for the accumulated SNR as a function of time by computing these statistics over the different realizations of the populations.
The solid lines indicate the median values, and the shaded regions indicate the symmetric 68\% confidence intervals.
Since the events in each population occur at different times, these are computed by first interpolating the SNR for each population as a function of time at evenly spaced points, and then computing these statistics at each of these points.

\subsection{Discussion of displacement and spin memory results}

\begin{figure}
  \includegraphics[width=\linewidth]{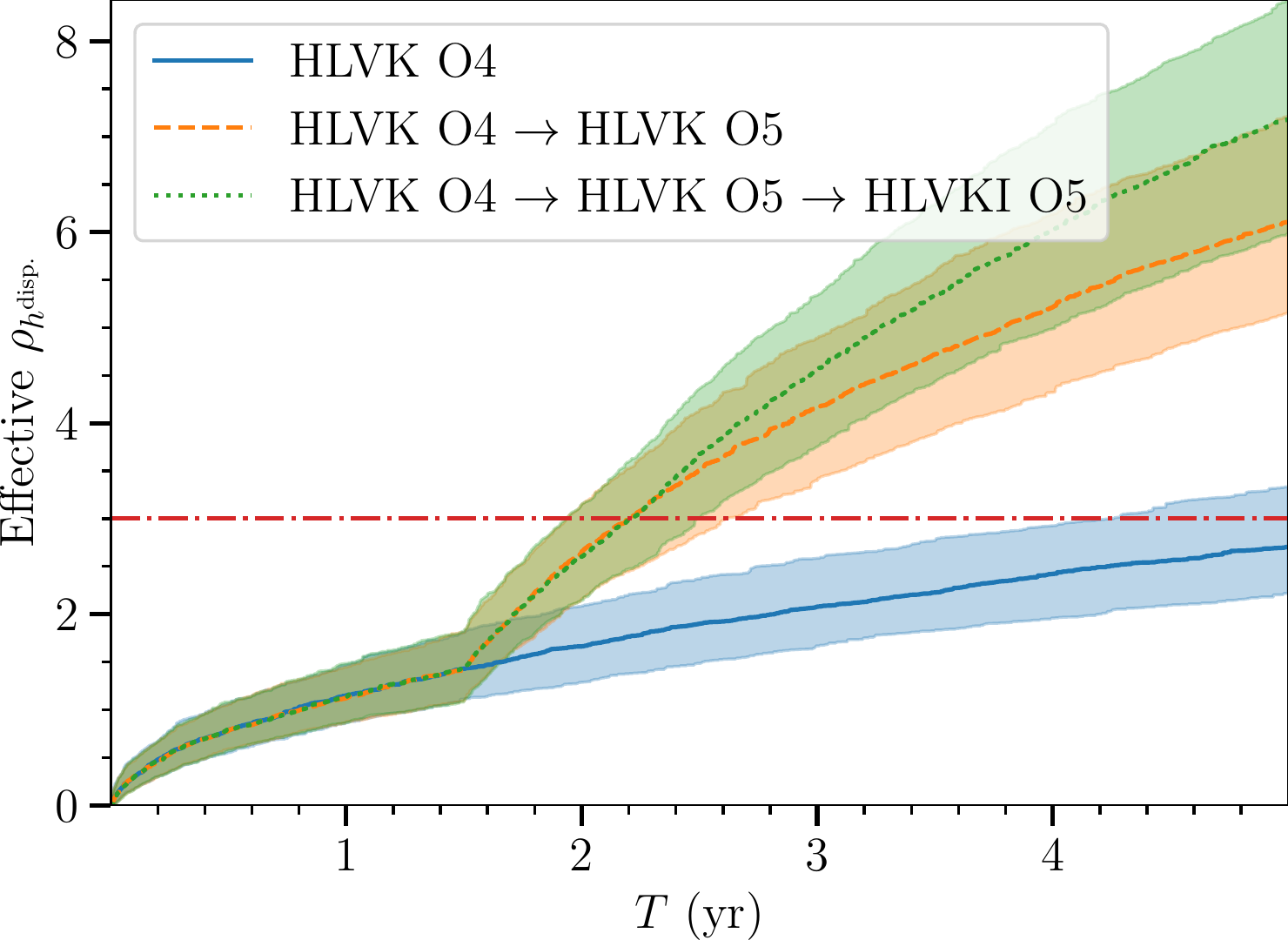}
  \caption{\label{fig:2g_disp} The median accumulated SNR and 68\% symmetric credible region for the displacement memory in second-generation detector networks as a function of run time for different realizations of binary-black-hole populations.
   The blue region corresponds to the O4 HLVK network at design sensitivity for all four detectors, which can be compared with the results of~\cite{Boersma2020}.
    The orange region instead accounts for an upgrade to the O5 sensitivity of the HLVKI network after $1.5$ years, as is currently planned~\cite{KAGRA2013}, and the green region accounts for the addition of LIGO India after $2.25$ years.
    The red dashed line is an SNR of 3, which corresponds via Eq.~\eqref{eqn:logBayes} to a log Bayes factor of 9, which would indicate strong evidence for the hypothesis that there is displacement memory in the population of black-hole mergers.}
\end{figure}

\begin{figure}
  \includegraphics[width=\linewidth]{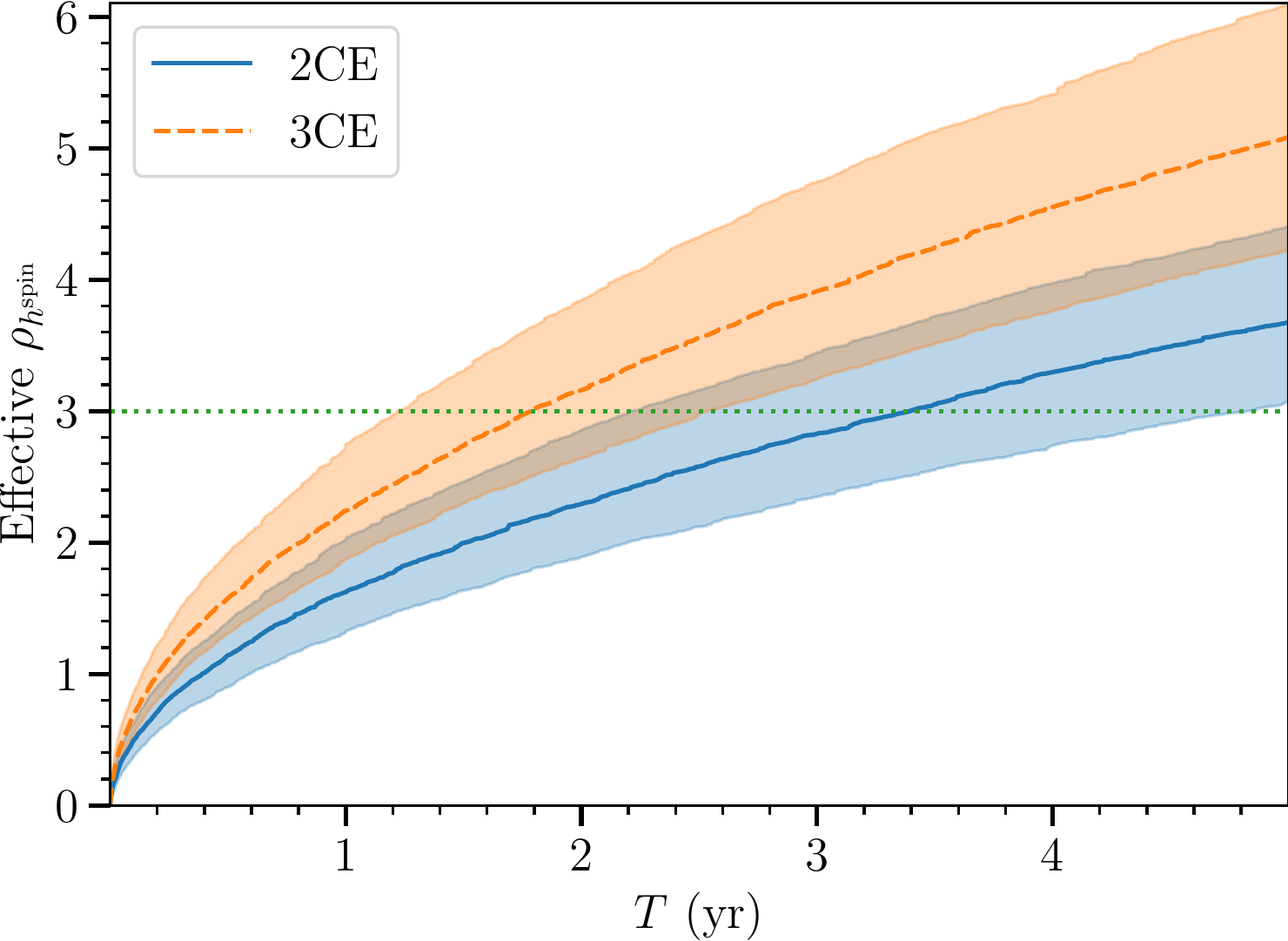}
  \caption{\label{fig:3g_spin} The median accumulated SNR and 68\% symmetric credible region for the spin memory in the third-generation Cosmic Explorer detectors as a function of run time for different realizations of binary-black-hole populations.
    The blue and orange regions correspond to two- and three-detector networks, respectively.
    The green dashed line corresponds to an accumulated SNR of 3 and is reached by the median of the populations for Cosmic Explorer for a three-detector network.}
\end{figure}

We can compare Fig.~\ref{fig:2g_disp} to Fig.~4 of~\cite{Boersma2020}, which used the O4 sensitivity for LIGO and Virgo (HLV) and the power-law population model from the first gravitational-wave transient catalog (GWTC-1).
The comparison illustrates that the population model from GWTC-2~\cite{LSC2020c} produces an estimate for the detection prospects of the gravitational wave memory effect that is somewhat less optimistic than that of the GWTC-1 populations.
By allowing the network to upgrade from O4 to O5 sensitivity, we find that detection prospects are more optimistic: after only $\sim 2.5$ years of total run time, the displacement memory will likely be measured with an accumulated SNR $\rho_{\rm eff.} = 3$, with the median reaching this threshold after $\sim 2.2$ years.
According to timeline in~\cite{KAGRA2013}, LIGO India will start $0.75$ years into O5, and so its contribution to the time of the initial detection of the memory is minimal.
However, after a total run time of $5$ years, including LIGO India will increase the accumulated SNR from $\rho_{\rm eff.} \simeq 6$ to $\rho_{\rm eff.} \simeq 7$, so it will help increase the significance of the detection.

From Fig.~\ref{fig:3g_spin}, we can similarly see that, after only two years of run time for the Cosmic Explorer network, the accumulated SNR in the spin memory signal will be just over $\rho_{\rm eff.} = 3$ for the median population, assuming that there are three Cosmic Explorer detectors in Europe, the United States, and Australia (3CE).
If there are only two detectors, the accumulated SNR in the spin memory will reach this threshold after $3.5$ years.
However, as we discussed in Sec.~\ref{sec:pop_model}, limiting to $z_{\rm max} = 1$ is very likely underestimating the effective SNR, and so the results will most likely be more optimistic than those illustrated in Fig.~\ref{fig:3g_spin}.
A feature that stands out in this plot is that the median and boundaries of the credible region appear smoother than those in Fig.~\ref{fig:2g_disp}.
The main reason for this is that many more events exceed the threshold of $\rho_{h_{\rm odd}}^2 > 2$ in Cosmic Explorer.
Events then are detected at a much higher rate, which leads to a shorter time between events and a smoother appearance of the curves.
This also implies that while measurements of displacement memory in second-generation detectors will involve a smaller number of the loudest events, the measurement of spin memory in third-generation detectors will be dominated by a large number of more average SNR events for the detector network.

\begin{figure}
  \includegraphics[width=\linewidth]{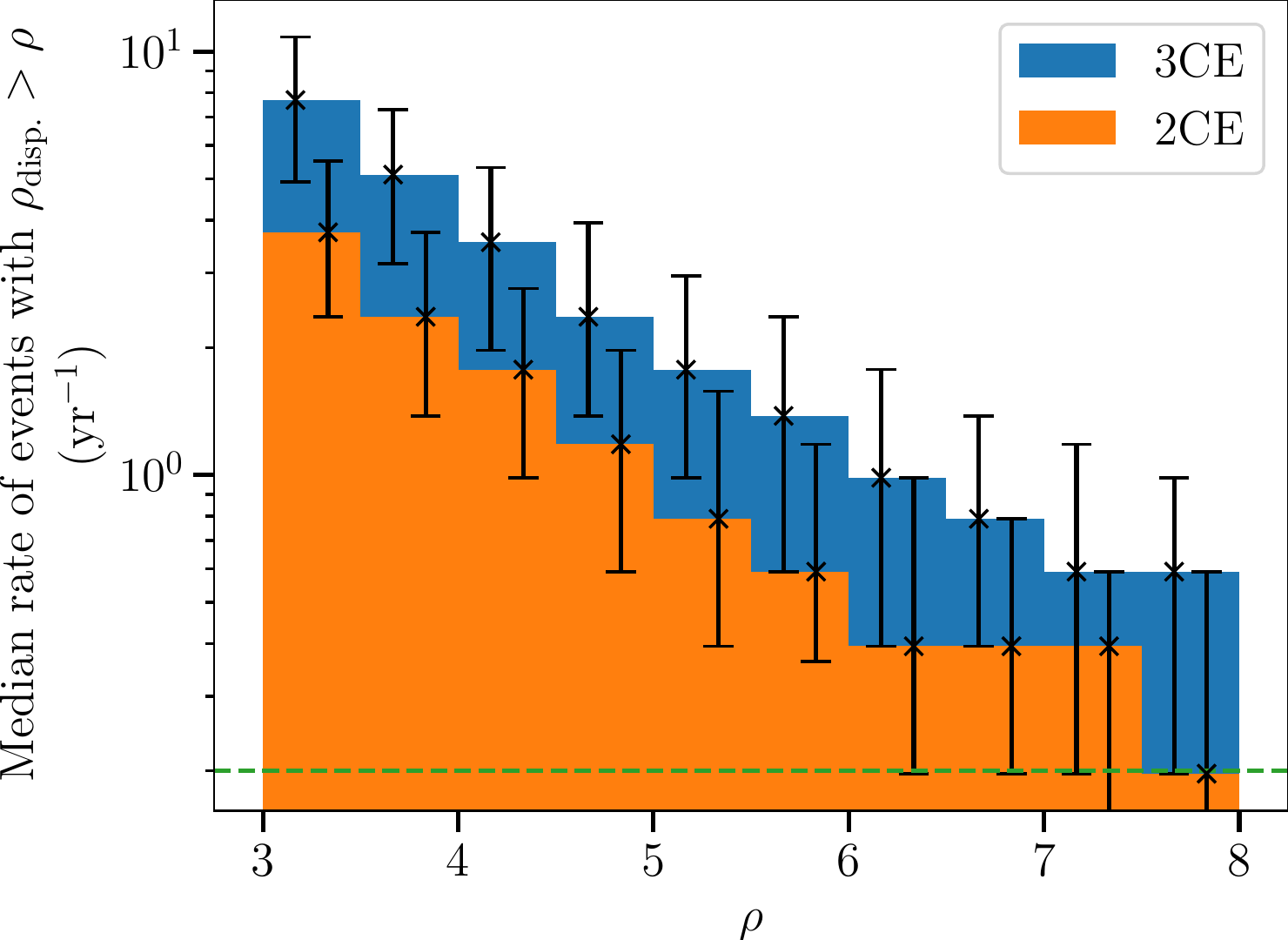}
  \caption{\label{fig:3g_disp} The median rate of events with an SNR in the displacement memory higher than some given amount, in Cosmic Explorer with either a two- (orange) or three- (blue) detector network.
    Error bars correspond to the symmetric 68\% confidence interval.
    The green line corresponds to seeing a single event over 5 years.}
\end{figure}

Finally, we show in Fig.~\ref{fig:3g_disp} a cumulative histogram of the median rate of individual events which have an SNR greater than 3 in the displacement memory for Cosmic Explorer.
This histogram is computed from $5$ years of data, and (to speed up the calculation) we restricted to $z_{\rm max} = 0.3$, as we did not find many loud events at higher redshifts.
For the median population, this histogram shows that two Cosmic Explorer detectors are expected to see around 3-4 events with an SNR greater than 3 per year, while three Cosmic Explorer detectors would 7-8.
These detectors may even see a handful of events with an SNR in the displacement memory greater than 7, over the span of 5 years.

\subsection{Dependence of the results on maximum redshift}

As we noted above in Sec.~\ref{sec:pop_model}, the maximum redshift to which Ref.~\cite{LSC2020c} recommended using its population model is $z_{\rm max} = 1$.
Because the SNR falls off with the inverse of the luminosity distance, but the number of events increases with the volume of space surveyed, it is not obvious \textit{a priori} whether the fewer louder and closer events contribute more to the SNR than the quieter and more distant events do (assuming these events pass the threshold for the SNR in the odd part of the waveform).
For second-generation detectors, empirical checks found that it was the nearest events that were most important.
For example,~\cite{Boersma2020} noted that it did not see much difference in the effective SNR when applying a distance cutoff of $2$ Gpc and when including more distant events.
Placing a cutoff in distance has the benefit of saving computational time by only considering the subset of events that contribute to the effective SNR.

\begin{figure}
  \includegraphics[width=\linewidth]{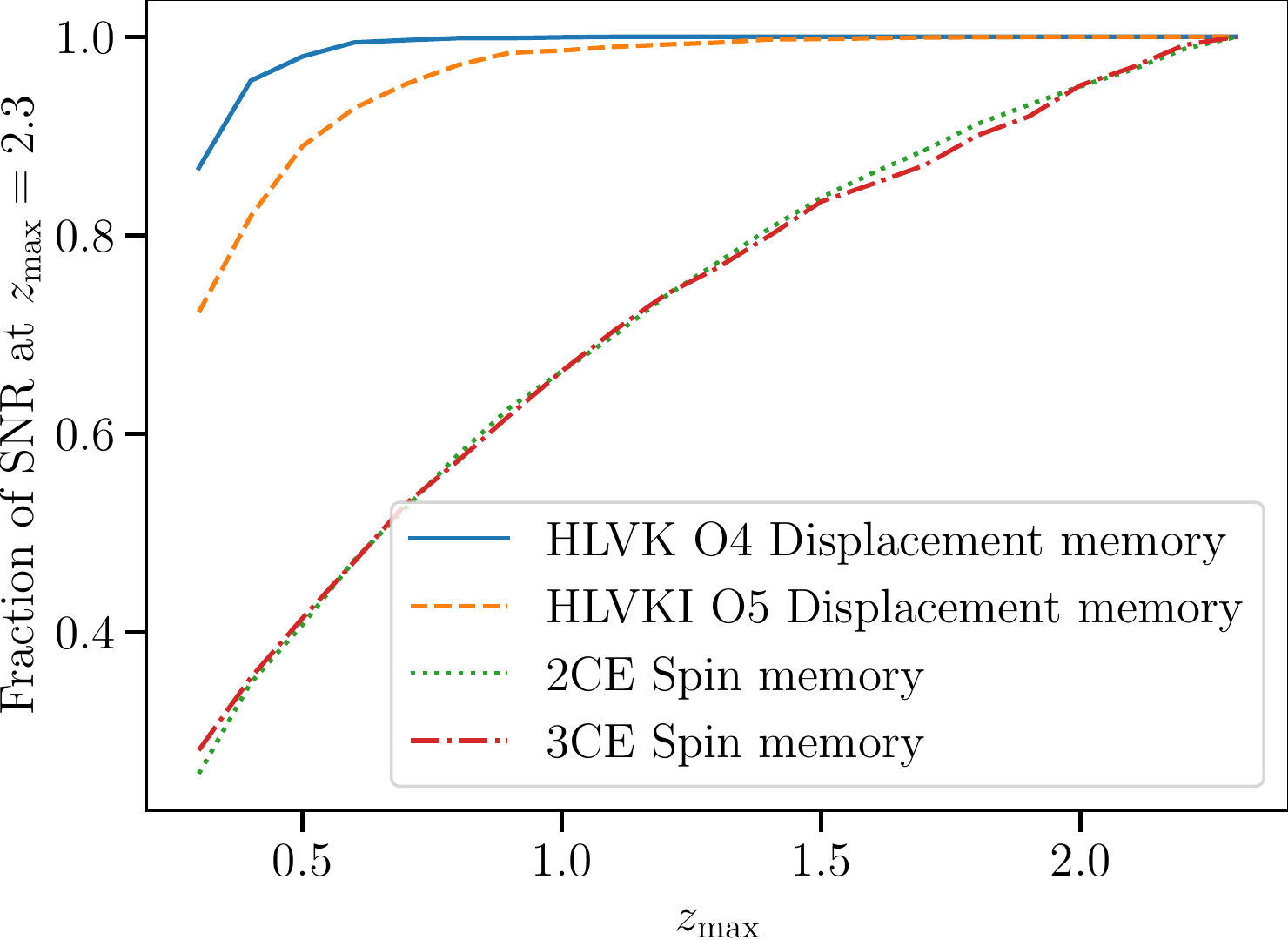}
  \caption{\label{fig:redshift} A plot of the final accumulated SNR after $0.1$ years for a given memory signal in a particular detector network, as a function of redshift cutoff $z_{\rm max}$, as a fraction of the SNR computed using a redshift cutoff of $z_{\rm max} = 2.3$.}
\end{figure}

To investigate the effects of a cutoff in redshift, in Fig.~\ref{fig:redshift} we plot the accumulated SNR of the displacement memory signal after $0.1$ years for different redshift cutoffs.
We show both second-generation detector sensitivities for O4 and O5, as well as the accumulated SNR of the spin memory signal after the same amount of time for Cosmic Explorer.
These plots show that a redshift of $0.8$ is sufficient for both O4 and O5, which we used in Fig.~\ref{fig:2g_disp}.
Note that this redshift, which corresponds to a luminosity distance cutoff of $\sim 5$ Gpc, is larger than that in~\cite{Boersma2020}, but comparable to that of~\cite{Hubner2019}.

For Cosmic Explorer, however, Fig.~\ref{fig:redshift} shows that we should \emph{not} use a redshift cutoff at all: even up to a redshift of $2.3$ (which is the maximum redshift at which an event could possibly have been seen so far~\cite{LSC2020c}), the accumulated SNR after $0.1$ years is still growing with increasing redshift.
Therefore, while we use the redshift cutoff of $z = 1$ for Fig.~\ref{fig:3g_spin}, this comes with the caveat that this is very likely underestimating the effective SNR.

\section{Conclusions and future directions} \label{sec:conclusions}

In this paper, we have estimated the detection prospects for the displacement and spin memory effects using simulated populations of binary black hole mergers based on the most favored model inferred from the detections in LIGO and Virgo's second gravitational-wave transient catalog.
Based on a detection criteria of an effective signal-to-noise ratio of 3 in a population, and similarly a signal-to-noise ratio of 3 for an individual event, we have found that
\begin{enumerate}

\item the LIGO, Virgo, and KAGRA detectors can detect the displacement memory in the population of binary mergers after 2.5 years when operating at design sensitivity (1.5 years) and ``plus'' sensitivity (1 year) (an observing scenario consistent with that in~\cite{KAGRA2013});

\item Cosmic Explorer can measure the spin memory effect in the median population of events if it operates for 2 years; and

\item Cosmic Explorer will, on average, have 7-8 events per year for which the displacement memory can be measured for a single event.

\end{enumerate}
The assumptions that go into these forecasts were enumerated throughout the text and will not be repeated here.

For the displacement memory, this work updates the results of~\cite{Lasky2016, Hubner2019, Boersma2020} to use populations consistent with the detections from the first two and a half observing runs~\cite{LSC2020c} and to account for the updated observing scenarios including the O5 observing run (upgrades to ``plus'' sensitivities).
Only updating the population model did not significantly affect the time needed to detect the memory, as our results are consistent with those of~\cite{Boersma2020} (the 68\% confidence intervals for our results overlap).
The shorter time to detection found in this paper over the five-year frame given in~\cite{Boersma2020} is, however, primarily determined by the increased sensitivity of the detectors during the O5 observing run.
We have also performed an initial check that using the \emph{most} recent distributions for population parameters from the third catalog~\cite{LSC2021b} does not make any significant difference from using just the second catalog~\cite{LSC2020c} as we did in this paper.
A more systematic comparison of the forecasts from different mass models (our forecasts for both only use the ``Power Law + Peak'' model) is certainly one possible future direction.
Another direction could be to refine the forecasts once O4 data is analyzed, since detection only becomes likely after the O5 run progresses for a few years.

The forecasts for the spin memory presented in this paper were new (as far as we are aware).
They were more computationally intensive, because of the large number of events detected by third-generation detectors, and, as a result, they could be refined in a few ways.
As with the displacement memory, further detections during O4 and O5 will improve our knowledge of the population, allowing for more accurate forecasts.
In particular, with the increased detection horizon during these runs, we expect to learn more about the population of distant (meaning $z > 1$) binaries, which are important for the spin memory---since the total SNR is dominated by a large number of events with small SNR in the spin memory---but are not well constrained by the current population models.
That these events contribute the most to the total SNR also yields two further issues: first, one may need to disentangle overlapping events, and second the spin memory may be small enough to be affected more by the detectors' calibration and gravitational waveform model uncertainties.
Gaining a better understanding of these potential systematic effects is beyond the scope of this initial study.

Finally, the displacement and spin memory effects are examples of so-called ``persistent observables''~\cite{Flanagan2019}, which are effects that a set of observers can measure by comparing measurements before and after a burst of gravitational waves.
These persistent observables in general arise from different nonoscillatory features of a gravitational wave signal which would manifest as lasting offsets in different numbers of time integrals of the gravitational-wave strain.
For example, observers with constant, nonzero relative acceleration would measure a persistent observable related to two time integrals of the gravitational-wave strain.
The simplest of these persistent observables, the ``curve deviation,'' arises from a lasting displacement measured by observers with an initial relative separation, velocity, and acceleration.
The piece of the lasting displacement that depends on initial separation is related to the displacement memory; the piece depending on initial relative velocity contains both the spin memory and the center-of-mass memory~\cite{Nichols2018}.
The pieces which depend on the initial acceleration (and higher derivatives such as the initial jerk), however, are distinct from these better-studied effects, and their related gravitational-wave signals have not yet been computed for binary mergers.
These higher-order observables were analyzed in general asymptotically flat spacetimes in~\cite{Grant2021}, and it was found that each higher-order effect probes distinct nonlinear aspects of the propagation of gravitational waves [each piece of it obeys equations similar to Eq.~\eqref{eqn:conservation}].
Once the phenomenology of these new observables is studied for compact binary sources in more detail, and in particular once a useful definition of a ``memory signal'' (as in Sec.~\ref{sec:mem_signal}) is established, we will investigate the detection prospects in future work.

\acknowledgments

D.A.N.\ acknowledges support from the NSF Grant No. PHY-2011784.
A.M.G.\ thanks Colm Talbot for clarifying remarks on the normalization of the PLPP primary mass model in~\cite{Talbot2018a} (see Footnote~\ref{fn:gauss_max}).
We also thank an anonymous referee for providing helpful feedback on an earlier draft of this work (see, in particular, Footnote~\ref{fn:sampling}).
We finally acknowledge Research Computing at The University of Virginia for providing computational resources and technical support that have contributed to the results reported within this publication.

\appendix

\section{Averages over Gaussian noise}

In this appendix, we compute averages of noise-weighted inner products over realizations of zero-mean Gaussian noise $n$.
We use them in Sec.~\ref{sec:stacking}, for example, to determine expected values of the Bayes factor.
Our main result is the following: for any real, deterministic function $a$, the expected value of a power of the noise-weighted inner product of $a$ with Gaussian noise $n$ is given by
\begin{equation} \label{eqn:avg_power}
  \expect[\<a|n\>^m] = \begin{cases}
    (m - 1)!! \<a|a\>^{m/2} & m \textrm{ even} \\
    0 & m \textrm{ odd}
  \end{cases},
\end{equation}
where the double factorial satisfies
\begin{equation} \label{eqn:double_fact}
  (2k - 1)!! = \frac{(2k)!}{2^k k!}.
\end{equation}
An immediate consequence of these two equations which we use throughout this paper is that
\begin{equation} \label{eqn:avg_exp}
  \begin{split}
    \expect[e^{\beta \<a|n\>}] &= \sum_{k = 0}^\infty \frac{(2k - 1)!!}{(2k)!} (\beta^2 \<a|a\>)^k \\
    &= \sum_{k = 0}^\infty \frac{1}{k!} \left(\frac{\beta^2}{2} \<a|a\>\right)^k = e^{\beta^2 \<a|a\>/2},
  \end{split}
\end{equation}
for any constant $\beta$.

In order to derive Eq.~\eqref{eqn:avg_power}, we only need to compute the expectation value of a product of the noise evaluated at different times.
This follows from the linearity of the noise-weighted inner product and the Fourier transform.
To compute this expectation value, we start by introducing the characteristic function $\Theta_m$ for a collection of $m$ random variables $\xi_1, \ldots, \xi_m$:
\begin{equation}
  \Theta_m (u_1, \ldots, u_m; \xi_1, \ldots, \xi_m) \equiv \expect\left[\exp\left(i \sum_{i = 1}^m u_i \xi_i\right)\right].
\end{equation}
This equation implies the following useful expression for the following expectation value:
\begin{equation} \label{eqn:m_point_fn}
  \begin{split}
    \expect[\xi_1 &\cdots \xi_m] \\
    &= \frac{1}{i^m} \left.\frac{\partial^m \Theta(u_1, \ldots, u_m; \xi_1, \ldots, \xi_m)}{\partial u_1 \cdots \partial u_m}\right|_{u_1 = \cdots = u_m = 0}.
  \end{split}
\end{equation}

The noise $n$ is a stationary Gaussian process with zero mean, and so the characteristic function for $n$ evaluated at the times $t_1, \ldots, t_m$ is given by (see, for example, Chapter 3 of~\cite{Stratonovich1963})
\begin{equation} \label{eqn:Theta_m}
  \begin{split}
    \Theta_m &[u_1, \ldots, u_m; n(t_1), \ldots, n(t_m)] \\
    &= \exp\left[-\frac{1}{2} \sum_{i, j = 1}^m u_i u_j k_2 (t_i - t_j; n)\right],
  \end{split}
\end{equation}
where $k_2$ is the correlation function of the noise.
The correlation function is related to the power spectral density $S_n(f)$ by a Fourier transform:
\begin{equation} \label{eqn:k_psd}
  \tilde k_2 (f; n) = \frac{1}{2} S_n (f).
\end{equation}
Note that $k_2$ and $\tilde k_2$ are even.
As Eq.~\eqref{eqn:Theta_m} has only even powers of each $u_i$ in its power series, it follows from Eq.~\eqref{eqn:m_point_fn} that
\begin{equation}
  \expect[n_1 (t_1) \cdots n_{2k + 1} (t_{2k + 1})] = 0.
\end{equation}
This immediately proves the odd $m$ case of Eq.~\eqref{eqn:avg_power}.

For the remainder of this appendix, we restrict to even $m$ and write $m = 2k$.
Next, the fact that in Eq.~\eqref{eqn:m_point_fn} the right-hand side is evaluated at $u_1 = \cdots = u_{2k} = 0$ implies that the contributing terms in this expression must have exactly $2k$ distinct $u_i$'s.
Thus, the collection of indices which appear must be a permutation of $\{1, \ldots, 2k\}$; we denote this set of permutations by $P_{2k}$.
Note, moreover, that each permutation occurs in the expansion of the exponential in Eq.~\eqref{eqn:Theta_m} exactly once, so we find
\begin{equation} \label{eqn:m_point_full_perm}
  \expect[n(t_1) \cdots n(t_{2k})] = \frac{1}{k! 2^k} \sum_{\pi \in P_{2k}} \prod_{i = 1}^k k_2 [t_{\pi(2i - 1)} - t_{\pi(2i)}; n],
\end{equation}
where the factor if $i^{2k} = (-1)^k$ cancels the same factor in the expansion of the exponential.
The notation $\pi(i)$ denotes the $i$\textsuperscript{th} element of the permutation $\pi$ of $\{1, \ldots, 2k\}$.
Now, note that certain of these permutations give exactly the same term in the sum: since $k_2$ is even, if two permutations $\pi$ and $\pi'$ are equal, except that
\begin{equation} \label{eqn:flip_pair}
  \pi(2i - 1) = \pi'(2i), \qquad \pi(2i) = \pi'(2i - 1),
\end{equation}
for all $i$ in some subset of $\{1, \ldots, k\}$, then these permutations yield the same contribution to the sum.
Similarly, if $\pi$ and $\pi'$ agree up to
\begin{equation} \label{eqn:permute_pairs}
  \pi(2i - 1) = \pi'(2j - 1), \qquad \pi(2i) = \pi'(2j),
\end{equation}
for all $(i, j)$ in some subset in $\{1, \ldots, k\} \times \{1, \ldots, k\}$, then these permutations give the same term in the sum, since they simply differ by the order in which each $k_2$ appears.
Since there are $2^k$ permutations which are related by Eq.~\eqref{eqn:flip_pair} and $k!$ permutations which are related by Eq.~\eqref{eqn:permute_pairs}, we can cancel the factor of $1/(2^k k!)$ in Eq.~\eqref{eqn:m_point_full_perm} by considering only a subset, denoted $\tilde P_{2k}$, containing permutations $\pi$ such that
\begin{equation}
  \pi(2i - 1) > \pi(2i),
\end{equation}
for all $i$, and
\begin{equation}
  \pi(1) > \pi(3) > \cdots > \pi(2k - 1).
\end{equation}
The final result is that
\begin{equation} \label{eqn:m_point_perm}
  \expect[n(t_1) \cdots n(t_{2k})] = \sum_{\pi \in \tilde P_{2k}} \prod_{i = 1}^k k_2 [t_{\pi(2i - 1)} - t_{\pi(2i)}; n].
\end{equation}
Denoting the number of elements in a set $S$ by $|S|$, note that, while $|P_{2k}| = (2k)!$, we find that Eq.~\eqref{eqn:double_fact} implies that
\begin{equation} \label{eqn:tilde_P_2k}
  |\tilde P_{2k}| = (2k - 1)!!.
\end{equation}

We next use Eq.~\eqref{eqn:m_point_perm} to compute the average of the Fourier transforms of the noise:
\begin{widetext}
\begin{equation}
  \expect[\tilde n(f_1) \cdots \tilde n(f_{2k})] = \sum_{\pi \in \tilde P_{2k}} \int \prod_{i = 1}^k \ud t_{\pi(2i - 1)} \ud t_{\pi(2i)} \exp\left\{2\pi i \left[f_{\pi(2i - 1)} t_{\pi(2i - 1)} + f_{\pi(2i)} t_{\pi(2i)}\right]\right\} k_2 [t_{\pi(2i - 1)} - t_{\pi(2i)}; n].
\end{equation}
Te simplify the integrals, we define the variables
\begin{equation}
  \tau_{\pi, i} \equiv t_{\pi(2i - 1)} - t_{\pi(2i)}, \qquad T_{\pi, i} \equiv \frac{1}{2} [t_{\pi(2i - 1)} + t_{\pi(2i)}],
\end{equation}
so that
\begin{equation}
  t_{\pi(2i - 1)} = T_{\pi, i} + \frac{1}{2} \tau_{\pi, i}, \qquad t_{\pi(2i)} = T_{\pi, i} - \frac{1}{2} \tau_{\pi, i} .
\end{equation}
The Jacobian determinant of the transformation is given by
\begin{equation}
  \begin{vmatrix}
    \dfrac{\partial t_{\pi(2i - 1)}}{\partial T_{\pi, i}} & \dfrac{\partial t_{\pi(2i - 1)}}{\partial \tau_{\pi, i}} \\
    \dfrac{\partial t_{\pi(2i)}}{\partial T_{\pi, i}} & \dfrac{\partial t_{\pi(2i)}}{\partial \tau_{\pi, i}}
  \end{vmatrix} = -1.
\end{equation}
We therefore have that
\begin{equation}
  \expect[\tilde n(f_1) \cdots \tilde n(f_{2k})] = \sum_{\pi \in \tilde P_{2k}} \int \prod_{i = 1}^k \ud \tau_{\pi, i} \ud T_{\pi, i} \exp\left\{\pi i \left[f_{\pi(2i - 1)} - f_{\pi(2i)}\right] \tau_{\pi, i} + 2\pi i \left[f_{\pi(2i - 1)} + f_{\pi(2i)}\right] T_{\pi, i}\right\} k_2 [\tau_{\pi, i}; n].
\end{equation}
Doing the integral over each $T_{\pi, i}$ gives delta functions $\delta[f_{\pi(2i - 1)} + f_{\pi(2i)}]$.
Because this will then require that $f_{\pi(2i - 1)} = -f_{\pi(2i)}$, we can impose this condition and perform the integral over each $\tau_{\pi, i}$.
This gives the Fourier transform of $k_2$, and the resulting expression can be written as
\begin{equation}
  \expect[\tilde n(f_1) \cdots \tilde n(f_{2k})] = \sum_{\pi \in \tilde P_{2k}} \prod_{i = 1}^k \delta[f_{\pi(2i - 1)} + f_{\pi(2i)}] \tilde k_2 [f_{\pi(2i - 1)}; n].
\end{equation}

With Eq.~\eqref{eqn:k_psd}, the noise-weighted inner product can be written as
\begin{equation}
  \<a|b\> = \int_{-\infty}^\infty \ud f \frac{\overline{\tilde a(f)} \tilde b(f)}{\tilde k_2 (f; n)},
\end{equation}
for real $a$ and $b$.
Therefore, assuming that $a_1, \ldots, a_{2k}$ are real and deterministic, we have that
\begin{equation}
  \begin{split}
    \expect[\<n|a_1\> \cdots \<n|a_{2k}\>] &= \int \ud^{2k} f \frac{\expect[\tilde n(-f_1) \cdots \tilde n(-f_{2k})] \tilde a_1 (f_1) \cdots \tilde a_{2k} (f_{2k})}{\tilde k_2 (f_1; n) \cdots \tilde k_2 (f_{2n}; n)} \\
    &= \sum_{\pi \in \tilde P_{2k}} \int \prod_{i = 1}^k \ud f_{\pi(2i - 1)} \ud f_{\pi(2i)} \frac{\delta(f_{\pi(2i - 1)} + f_{\pi(2i)}) \tilde a_{\pi(2i - 1)} (f_{\pi(2i - 1)}) \tilde a_{\pi(2i)} (f_{\pi(2i)})}{\tilde k_2 [f_{\pi(2i)}; n]} \\
    &= \sum_{\pi \in \tilde P_{2k}} \prod_{i = 1}^k \<a_{\pi(2i - 1)}|a_{\pi(2i)}\>,
  \end{split}
\end{equation}
\end{widetext}
where we have used the fact that $\overline{\tilde a_i (f)} = a_i (-f)$ and $\tilde k_2 (f; n) = \tilde k_2 (-f; n)$.
Finally, using Eq.~\eqref{eqn:tilde_P_2k} and setting $a_1 = \cdots = a_{2k} = a$, the even case of Eq.~\eqref{eqn:avg_power} follows.

\section{Unknown sign of the memory} \label{app:sign}

In the analysis in this paper, we ignored the events for which the sign of the memory could not be determined (specifically, by rejecting events for which the SNR$^2$ in the odd part of the oscillatory waveform was less than two in the detector network).
However, the analysis in~\cite{Hubner2019} did not ignore such events, and it is argued there that such events contribute much less to the overall Bayes factor (or ``effective'' SNR) in a population of events.
In this appendix, we provide an argument for why this is the case by using the same toy model which we considered in Sec.~\ref{sec:stacking}.

We begin our calculation of the Bayes factor by starting with Eq.~\eqref{eqn:bayes_factor_post}, but instead of using a posterior for a model without memory that is peaked around the true value of the parameters $\theta_0$, we use
\begin{equation}
  p[h_{\rm osc} (\theta) | d] = (1 - \alpha) \delta(\theta - \theta_0) + \alpha \delta(\theta - \theta_0').
\end{equation}
As in Sec.~\ref{sec:sign}, $\theta_0'$ are the approximately degenerate values of the parameters that give the incorrect sign of the memory.
The factor of $\alpha$ in the posterior $p[h_{\rm osc} (\theta)|d]$ describes how well the degeneracy is broken; $\alpha = 0$ is the case where the degeneracy is completely broken, and $\alpha = 1/2$ is the case were it is completely unbroken.
Substituting this posterior into the expression for the Bayes factor in Eq.~\eqref{eqn:bayes_factor_post} we obtain after some calculation
\begin{widetext}
\begin{equation}
  \mc B^{\rm mem.}_{\rm no\,mem.} (d) = (1 - \alpha) \exp\left[\frac{1}{2} \rho_{h_{\rm mem.} (\theta_0)}^2 + \<h_{\rm mem.} (\theta_0)|n\>\right] + \alpha \exp\left[\frac{1}{2} \rho_{h_{\rm mem.} (\theta_0')}^2 + \<h_{\rm mem.} (\theta_0') | 2h_{\rm odd} (\theta_0) + n\>\right].
\end{equation}
Note that $h_{\rm odd}$ in the expression above includes the odd parts of both the oscillatory and memory signals.
Using Eq.~\eqref{eqn:avg_exp} and assuming, for simplicity, the same value of $\alpha$ for all noise realizations, we find after some additional calculations that
\begin{equation}
  \expect[\mc B^{\rm mem.}_{\rm no\,mem.} (d)] = (1 - \alpha) \exp\left[\rho_{h_{\rm mem.} (\theta_0)}^2\right] + \alpha \exp\left\{-\rho_{h_{\rm mem.} (\theta_0)}^2 + 2 [\<h_{\rm mem.} (\theta_0) | h_{\rm mem.}^{\rm even} (\theta_0)\> + \<h_{\rm mem.} (\theta_0') | h_{\rm osc.}^{\rm odd} (\theta_0)\>]\right\}.
\end{equation}
\end{widetext}

From this general expression, it is not clear whether, on average, the memory hypothesis is favored (that is $\expect[\mc B^{\rm mem.}_{\rm no\,mem.} (d)] \geq 1$).
However, we now argue that we can reasonably neglect the last two terms in the second exponential.
The first of these terms we expect to be smaller in magnitude than $\rho_{h_{\rm mem.} (\theta_0)}^2$ since the even part of the memory signal is subdominant.
The second term is also likely small, as the subdominant oscillatory and the nonoscillatory parts of the signal are not morphologically similar and would thus not have a significant overlap.
When we neglect these terms, we find that
\begin{equation} \label{eqn:bayes_estimate}
  \expect[\mc B^{\rm mem.}_{\rm no\,mem.} (d)] = \cosh_\alpha \rho_{h_{\rm mem.} (\theta_0)}^2,
\end{equation}
where we have defined
\begin{subequations}
  \begin{align}
    \cosh_\alpha x &\equiv (1 - \alpha) e^x + \alpha e^{-x}, \\
    \sinh_\alpha x &\equiv (1 - \alpha) e^x - \alpha e^{-x},
  \end{align}
\end{subequations}
as generalizations of the hyperbolic trigonometric functions $\cosh \xi$ and $\sinh \xi$ (which are $\cosh_{1/2} x$ and $\sinh_{1/2} x$, respectively).
In what follows, we simply assume that Eq.~\eqref{eqn:bayes_estimate} holds, keeping in mind that this is just an estimate for the average Bayes factor.
This estimate holds exactly when $h_{\rm osc.}$ is purely even and $h_{\rm mem.}$ is purely odd (for example, for nonprecessing binaries), as the two terms which we have neglected are linear in $h_{\rm osc.}^{\rm odd}$ and $h_{\rm mem.}^{\rm even}$, respectively; in this case, since $h_{\rm osc.}$ is purely even, the degeneracy is also completely unbroken (namely, $\alpha = 1/2$).

Given Eq.~\eqref{eqn:bayes_estimate}, we now investigate the bounds on $\expect[\mc B^{\rm mem.}_{\rm no\,mem.} (d)]$.
First, one can show that $\cosh_\alpha x$ and $\sinh_\alpha x$ satisfy the identities
\begin{gather}
  \cosh_\alpha^2 x - \sinh_\alpha^2 x = 4 (1 - \alpha) \alpha, \\
  \frac{\ud}{\ud x} \cosh_\alpha x = \sinh_\alpha x, \quad \frac{\ud}{\ud x} \sinh_\alpha x = \cosh_\alpha x .
\end{gather}
These equations can be used to show that
\begin{equation} \label{eqn:cosh_geq_1}
  \cosh_\alpha x \geq 1
\end{equation}
if $x \geq 0$ and $\alpha \leq 1/2$.
Because the SNR is not imaginary and because we do not consider systematic biases that could make the posterior favor the parameters $\theta'_0$, this will be sufficient for our calculations.
To prove Eq.~\eqref{eqn:cosh_geq_1}, note that
\begin{equation}
  \cosh_\alpha 0 = 1, \qquad \sinh_\alpha 0 = 1 - 2 \alpha \geq 0,
\end{equation}
and the only root of $\sinh_\alpha x$ is at $\ln \sqrt{\alpha/(1 - \alpha)}$, which is not positive when $\alpha \leq 1/2$.
As a result, if follows that $\sinh_\alpha x \geq 0$ for $x \geq 0$ and that $\cosh_\alpha x$ is increasing, which proves Eq.~\eqref{eqn:cosh_geq_1}.
Using Eq.~\eqref{eqn:bayes_estimate}, we therefore find that
\begin{equation}
  1 \leq \expect[\mc B^{\rm mem.}_{\rm no\,mem.} (d)].
\end{equation}
This lower bound implies that, even though the degeneracy is not broken, on average this signal will still increase the evidence for memory in the population.

We now discuss an upper bound on $\expect[\mc B^{\rm mem.}_{\rm no\,mem.} (d)]$.
First, note that
\begin{subequations} \label{eqn:ln_cosh_derivs}
  \begin{align}
    \frac{\ud}{\ud x} \ln \cosh_\alpha x &= \tanh_\alpha x, \\
    \frac{\ud^2}{\ud x^2} \ln \cosh_\alpha x &= 4 \alpha (1 - \alpha) \sech_\alpha^2 x \nonumber \\
    &\leq 4 \alpha (1 - \alpha) ,
  \end{align}
\end{subequations}
where $\tanh_\alpha x = \sinh_\alpha x/\cosh_\alpha x$ and $\sech_\alpha x = 1/\cosh_\alpha x$.
Using Taylor's theorem (see, for example, Theorem~5.15 of~\cite{Rudin1976}), which states that
\begin{equation}
  f(x) = \sum_{n = 0}^N \frac{f^{(n)} (a)}{n!} (x - a)^n + R_N (x),
\end{equation}
where
\begin{equation}
  R_N (x) = \frac{f^{N + 1} (x_L)}{(N + 1)!} (x - a)^{N + 1},
\end{equation}
for some $x_L \in [a, x]$, we can use Eq.~\eqref{eqn:ln_cosh_derivs} to give bounds on $\ln \cosh_\alpha x$:
\begin{equation}
  \ln \cosh_\alpha x \leq (1 - 2 \alpha) x + 2 \alpha (1 - \alpha) x^2.
\end{equation}
Applying the estimate in Eq.~\eqref{eqn:bayes_estimate} then yields
\begin{equation}
  \begin{split}
    \expect[\mc B^{\rm mem.}_{\rm no\,mem.} (d)] \leq \exp\Big[&(1 - 2 \alpha) \rho_{h_{\rm mem.} (\theta_0)}^2 \\
    &+ 2 \alpha (1 - \alpha) \rho_{h_{\rm mem.} (\theta_0)}^4\Big].
  \end{split}
\end{equation}
The upper bound shows that the expected value of the Bayes factor grows more slowly than in the case where the degeneracy is completely broken ($\alpha=0$).
When the degeneracy is completely unbroken ($\alpha = 1/2$), the Bayes factor's log grows with the SNR of the memory to the fourth power, as stated in~\cite{Hubner2019}.
For SNRs much less than one, this grows \emph{much} more slowly than the case where the sign of the memory is known.
However, for $\alpha$ even slightly above $1/2$, the growth becomes much faster due to its dependence on the square, instead of the fourth power, of the SNR.

Although this argument suggests that on average it would be beneficial to include all events and not just those for which the degeneracy is broken with strong confidence, for any individual noise realization, the Bayes factor could be less than one.
To aid with comparisons to prior literature~\cite{Lasky2016,Boersma2020} and to help make the forecasts more computationally efficient, we use only the events that satisfy the criteria described in Sec.~\ref{subsubsec:criteria}.

\section{Another degeneracy of the dominant quadrupolar mode} \label{app:degeneracy}

In this appendix, we discuss another transformation, separate from that in Eq.~\eqref{eqn:m_degen}, that is considered in Sec.~V.A.3 of~\cite{Nichols2017}:
\begin{equation} \label{eqn:iota_transform}
  \iota \to \pi - \iota.
\end{equation}
Since the spin-weighted spherical harmonics transform as
\begin{equation}
  \pb{s} Y_{lm} (\pi - \iota, \phi_{\rm ref.}) = (-1)^{l + s} \overline{\pb{s} Y_{l(-m)} (\iota, \phi_{\rm ref.})},
\end{equation}
it was noted in Ref.~\cite{Nichols2017} that the spin-memory mode would also flip sign under this transformation.
Thus, it was mentioned in~\cite{Nichols2017} that one would need to additionally need to distinguish the binary's inclination between $\iota$ and $\pi-\iota$ to stack the SNR for the spin memory mode to compute the effective SNR.
While this is true, the transformation~\eqref{eqn:iota_transform} is not a degeneracy of the dominant quadrupole mode, so distinguishing between $\iota$ and $\pi - \iota$ will not present an additional challenge.
This can be shown by computing the transformation of $h_{(lm)}$ under Eq.~\eqref{eqn:iota_transform}.
The result that we find is
\begin{equation} \label{eqn:l_degen}
  \begin{split}
    \sum_{|m| \leq l} h_{(lm)} \to (-1)^l \sum_{|m| \leq l} \Big\{&F_+ \Re\Big[\overline{h_{l(-m)}} \pb{-2} Y_{lm}\Big] \\
    &+ F_\times \Im\Big[\overline{h_{l(-m)}} \pb{-2} Y_{lm}\Big]\Big\}.
  \end{split}
\end{equation}
Note that, relative to Eq.~\eqref{eqn:antenna_modes}, there is a sign difference in the second line coming from $\Im[\bar z] = -\Im[z]$.
This transformation therefore treats plus and cross polarizations differently, with the cross polarization gaining an additional relative sign.
Using Eq.~\eqref{eqn:mem_symm}, and the fact that (in the quadrupole approximation) the displacement memory is purely plus polarized and the spin memory purely cross polarized, we therefore find that this transformation is a degeneracy of the displacement memory and flips the sign of the spin memory.
This transformation, however, is \emph{not} a degeneracy of the oscillatory part of the signal, as this part of the signal has nonzero plus- and cross-polarized components.
Thus, it does not result in the same ``sign-of-the-memory'' issue as the transformation in Eq.~\eqref{eqn:m_degen}.

However, if in addition to Eq.~\eqref{eqn:iota_transform}, one were to simultaneously perform the transformation\footnote{Since this transformation changes the location of the binary on the sky, any degeneracy that it induces could be broken by measuring the arrival times of the signal at the different detectors.}
\begin{equation}
  \psi \to -\psi, \quad \delta \to -\delta, \quad \alpha \to \alpha + \pi,
\end{equation}
this induces a transformation
\begin{equation}
  \bs X \to -\bs X, \qquad \bs Y \to \bs Y,
\end{equation}
so that the antenna patterns undergo $F_+ \to F_+$, $F_\times \to -F_\times$.
This changes the sign of the second line of Eq.~\eqref{eqn:l_degen}, which implies that that this transformation now is effectively the same as
\begin{equation}
  h_{lm} \to (-1)^l \overline{h_{l(-m)}}.
\end{equation}
However, by Eqs.~\eqref{eqn:osc_symm} and~\eqref{eqn:mem_symm}, we find that $h^{\rm osc.}$ (for even $l$), $h^{\rm disp.}$, and even $h^{\rm spin}$ are \emph{all} even under this transformation.
Therefore, once again, this transformation does not result in any sign-of-the-memory issue.

\bibliography{refs}

\end{document}